\documentclass[twoside]{article}

\usepackage[accepted]{aistats2025}
\usepackage{makecell}
\usepackage{lastpage}
\usepackage{fancyhdr}
\usepackage{amsmath}
\usepackage{amssymb}
\usepackage{graphicx,psfrag,epsf}
\usepackage{enumerate}

\usepackage{url} 
\usepackage{float}
\usepackage[table,xcdraw,dvipsnames]{xcolor}

\usepackage{cite}  

\usepackage{multirow}
\usepackage{lipsum}
\usepackage{amsfonts}
\usepackage{cite}      
                  
\usepackage{graphicx}   

\usepackage{amsthm}
\RequirePackage[colorlinks,citecolor=blue,urlcolor=blue]{hyperref}
\usepackage{amssymb,calc}
\usepackage{thmtools}
\usepackage{mathrsfs}
\usepackage{mathtools}
\usepackage{bm}
\usepackage{bbm}
\usepackage{caption}
\usepackage{subcaption}

\usepackage{enumerate}
\usepackage{rotating}
\RequirePackage{cleveref}
\usepackage{hyphenat}
\usepackage{natbib}

\usepackage{caption,subcaption}
\usepackage{tikz}
\usepackage{pgfplots}

\usepackage[T1]{fontenc}
\usepackage{textcomp} 
\usepackage{lmodern}

\usepackage{algorithm}
\usepackage[noend]{algpseudocode}
\makeatletter
\renewcommand{\ALG@name}{Algorithm}
\makeatother

\pgfplotsset{width=10cm}

\tikzset{declare function={gamma(\x)=sqrt(2*pi)*\x^(\x-0.5)*exp(-\x)*exp(1/(12*\x));}}
\tikzset{declare function={tpdf(\x,\nu)=gamma(0.5*(\nu+1))/(sqrt(pi*\nu)*gamma(\nu/2))*(1+\x^2/\nu)^(-(\nu+1)/2);}}
\tikzset{declare function={invgampdf(\x,\a,\b)=(\b/\x)^\a/\x/gamma(\a)*exp(-\b/\x);}}

\newcommand{\nhphantom}[1]{\ifmmode\settowidth{\dimen0}{$#1$}\else\settowidth{\dimen0}{#1}\fi\hspace*{-\dimen0}}

\def\C {\,|\:}

\newcommand\E{\mathbb E}

\renewcommand\d{\mathrm d}

\newcommand\R{\mathbb R}
\renewcommand\b{\bm{\beta}}

\newcommand\N{\mathbb N}

\newcommand{\indep}{\mathrel{\perp\!\!\!\!\perp}}
\newcommand*{\diff}{\mathop{}\!\mathrm{d}}

\newtheorem{theorem}{Theorem}
\newtheorem{assumption}{Assumption}
\newtheorem{lemma}[theorem]{Lemma}
\newtheorem{condition}{Condition}
\newtheorem{corollary}[theorem]{Corollary}

\renewcommand{\nhphantom}[1]{\ifmmode\settowidth{\dimen0}{$#1$}\else\settowidth{\dimen0}{#1}\fi\hspace*{-\dimen0}}

\numberwithin{equation}{section}

\crefname{thm}{Theorem}{Theorems}
\crefname{prop}{Proposition}{Propositions}
\crefname{lem}{Lemma}{Lemmas}
\crefname{coro}{Corollary}{Corollaries}
\crefname{add}{Addendum}{Addendums}
\crefname{asm}{Assumption}{Assumptions}
\crefname{alg}{Algorithm}{Algorithms}
\crefname{proc}{Procedure}{Procedures}
\crefname{exe}{Exercise}{Exercises}
\crefname{exa}{Example}{Examples}
\crefname{prob}{Problem}{Problems}
\crefname{section}{Section}{Sections}
\crefname{subsection}{Section}{Sections}
\crefname{appendix}{Appendix}{Appendices}

\DeclareMathOperator{\unif}{Unif}

\def\argmin{\mathop{\arg\min}}
\def\argsup{\mathop{\arg\sup}}
\def\arginf{\mathop{\arg\inf}}

\allowdisplaybreaks[3]

\usepackage{graphicx} 

\usepackage{algorithm}
\usepackage{algpseudocode}
\usepackage[table]{xcolor}

\usepackage{listings}

\lstdefinestyle{pythonstyle}{
    language=Python,
    basicstyle=\ttfamily\small,
    keywordstyle=\color{blue},
    stringstyle=\color{red},
    commentstyle=\color{green!60!black},
    morekeywords={nn, torch},
    breaklines=true,
    showstringspaces=false,
    frame=single,  
    numbers=none   
}

\begin{document}

\twocolumn[
\aistatstitle{Deep Generative Quantile Bayes}
\aistatsauthor{ Jungeum Kim \And Percy S. Zhai \And  Veronika Ro\v{c}kov\'{a}}
\aistatsaddress{The University of Chicago Booth School of Business} 
]

\begin{abstract}
We develop a   multivariate posterior sampling procedure through deep generative quantile learning.
Simulation proceeds implicitly through a  push-forward mapping that can transform   i.i.d. random vectors  samples from the posterior.
We utilize Monge-Kantorovich depth in multivariate quantiles to directly sample from Bayesian credible sets, a unique feature not offered by typical posterior sampling methods.
To enhance training of the quantile mapping, we design a neural network that automatically performs summary statistic extraction.
This additional neural network structure has performance benefits including  support shrinkage (i.e.  contraction of our posterior approximation) as the observation sample size increases. We demonstrate the usefulness of our approach on several examples where the absence of likelihood renders classical MCMC infeasible.
Finally, we provide the following frequentist theoretical justifications for our quantile learning framework:
{consistency of the estimated vector quantile, of  the recovered posterior distribution, and of the corresponding Bayesian credible sets.
}
\end{abstract}

\section{Introduction}
The purpose of this work is to develop a generative sampler from a Bayesian posterior for implicit models whose likelihoods can be accessed only through simulation.
We develop  a new approach based on quantile learning as an alternative   to existing adversarial samplers \citep{wang2022adversarial}. 
Outside Bayes, quantile learning has been useful across a broad spectrum of practical applications, particularly in contexts where the target distribution exhibits skewness or heavy tails, or when the tail behavior is of primary concern \citep{yu2003quantile}.
Recently, there has been a growing interest inside the statistical community in the application of quantile learning to generative modeling, from both Bayesian \citep{polson2023generative} and frequentist perspectives \citep{wang2024generative}.
Our work extends this focus from a one-dimensional to a multi-dimensional regime.

Defining a multivariate quantile presents a challenge due to the non-uniqueness of the mapping from a uniform distribution to the target multivariate distribution. Additionally, while the monotonicity of quantile functions can be ensured for univariate variables, this property does not automatically extend to the multivariate context. These ambiguities can be resolved by considering only those mappings that are a gradient of a convex potential function \citep{carlier2016vector}.

Following the approach of \cite{wang2022adversarial}, we obviate the need for MCMC by training our sampler on simulated data  obtained from a likelihood simulator (i.e., a forward sampler) and a prior simulator.
{
However, we take a different approach that learns the quantile mapping directly.
}
The training dataset consists of { generated} triplets $\{\theta_i, X_i, U_i\}_{i=1}^N$, where {$d$-dimensional} $\theta_i$'s are simulated from the prior $\pi(\theta)$ in the domain $\Theta\subset\R^d$, $X_i$'s are simulated from the implicit likelihood $L(\cdot \mid \theta)$ on the domain $\mathcal{X}\subset \R^{d_X\times n}$, and $U_i\sim F_U$ are i.i.d. random vectors from a source distribution $F_U$ on $\mathcal{U}\subset\R^d$. We train a conditional deep learning mapping $\hat Q_{\theta\mid X}: \mathcal{U} \rightarrow \Theta$ that aims to simulate from the posterior distribution $\pi(\theta\C X)$ by pushing  spherically uniform random vectors through itself. 
In the univariate case with $d=1$ and uniform $F_U$, this strategy corresponds to inverse transform sampling using an estimated quantile function   $\hat Q_{\theta\mid X}(\cdot)$.
{
The approach by \cite{polson2023generative} performs supervised learning on the triplets using the pinball loss.
This method does not simply generalize to the multivariate setting, where the relationship between $X_i$'s and $\theta_i$'s is more nuanced and cannot be simply aligned through the pinball loss function.
}
Instead, we aim to compute the 2-Wasserstein distance between the uniform distribution and the conditional distribution of $\theta$ given $X$, which naturally yields a transport map that can be used for posterior sampling.
This approach is different from \cite{wang2022adversarial} who iteratively estimate and minimize this Wasserstein distance {(See, Section \ref{sec:diff} for more detailed comparisons).}

{
A particular contribution of this paper is to incorporate summary statistics into posterior quantile maps.
Motivated by the Noise Outsourcing Lemma, this key technical extension not only brings feasibility to learning the convex potential function, but also enables a broad range of quantile learning methods with theoretical guarantee on their consistency.
The vital step of summary statistics learning has been broadly studied in the literature.
Long Short-Term Memory (LSTM, \cite{hochreiter1997long}) neural networks, for instance, handles correlated observations and is suitable when the order of the data points matters.
Meanwhile, the Deep Sets Neural Network (henceforth DeepSet, \cite{zaheer2017deep}) is designed for representing summary statistics for exchangeable data.
Our approach integrates both architectures for enhanced summary statistic learning.
}


{
Since the true posterior shrinks with the increase of dimensionality, the consistent credible sets should also shrink as $n$ increases, a phenomenon that we call \emph{support shrinkage}.
We empirically demonstrate that our method that applies DeepSet exhibits support shrinkage.
In addition, in the simulation study where the true posterior is known, the credible sets from our proposed method are close to the oracle sets even when the dimension $n$ is high, which agrees with our theoretical findings.
}
While \cite{jiang2017learning} also utilize deep learning for automatic summary statistic learning, their approach is more closely aligned with \cite{polson2023generative}, as they explicitly apply supervised learning to predict $\theta_i$ given $X_i$.

{
Our approach learns a push-forward map from a spherical uniform distribution.
Therefore, a credible set of arbitrary level $\tau\in(0,1)$ can be obtained by applying the map on the inner ball of radius $\tau$.
Compared to traditional Bayesian posterior sampling methods like MCMC or ABC that sample indirectly from posterior draws, no resampling is needed for our approach.
}
A formal definition of the credible set relies on the notion of data depth \citep{hallin2021distribution}.
Our choice is Monge-Kantorovich Depth \citep{chernozhukov2017monge}, which
can be viewed as a byproduct of  the vector quantile, interpreted as a potential function in the quantile space.
The equipotential surfaces  play a role of quantile contours, which can be equivalently regarded as  credible sets.


Numerous studies in the literature have explored the theory of deep quantile regression.
\cite{white1992nonparametric} used the method of sieves to establish the consistency of nonparametric conditional quantile estimators based on single hidden layer feed-forward networks.
\cite{padilla2022quantile} demonstrated consistency results of conditional quantile estimate that minimizes the pinball loss.
We build upon a more general framework by \cite{chernozhukov2017monge}, and {
demonstrate the asymptotic consistency of the estimated vector quantile.
We also prove that the recovered posterior of $\theta$ converges to the true posterior in terms of 2-Wasserstein distance.
}


Our contributions can be summarized as follows:
\begin{enumerate}
    \item We extend the approach of \cite{polson2023generative} from a one-dimensional to a $d$-dimensional parameter $\theta$ using two strategies. The first naive strategy consists of learning $d$ univariate samplers exploiting a chain-rule representation of the joint distribution $\pi(\theta\C X)$. Given a specific ordering of the variables in $\theta$, we learn these samplers sequentially by adding previous parameters (simulated from the previous univariate posterior samplers) into the training data table for the next parameter in the sequence {(See, Section \ref{sec:autoreg})}. Next, we develop our quantile learning approach { for generative Bayes}. 
     \item As a byproduct, our multivariate quantile learning method enables direct simulation from  {\em multivariate} Bayesian credible sets. Credible sets are  fundamental  for Bayesian inference and we can target them directly without imposing any strict geometry structure (which would otherwise be the case using an MCMC or ABC approach). {A convex hull of the sampled points then provides an estimate of the credible set}.
     \item  Not all deep learning architectures can be equally useful in generative modeling. We design a particular network for automatic summary statistics learning capable of handling both an increasing number of observations and the dependencies between them.
     Using this approach, we observe contraction of estimated credible sets with increasing sample size, a phenomenon  that we call {\em support shrinkage}.
     {
     If the credible sets are converging to the oracle sets based on the true posterior, support shrinkage is a necessary sign.
     }
     
     \item We provide frequentist theory for our multivariate quantile learning  approach as well as the initial approach of \cite{polson2023generative}.
     Existing theoretical results on unidimensional quantile learning cannot be directly extended to the multidimensional case; our work addresses this gap.
     {Specifically, we demonstrate that as $N\rightarrow\infty$ (1) the estimated vector quantile function achieves uniform consistency, (2) the recovered posterior converges uniformly to the true posterior in terms of 2-Wasserstein distance, and (3) the Bayesian credible sets converge to oracle sets.}
     {
     These generic theoretical results apply to all quantile learning methods that learn the convex potential function based on the summary statistics using a feed-forward neural network.
     }
\end{enumerate} 

The rest of the paper is outlined as follows. In Section \ref{sec:multi_q}, we review the recent advances in multi-dimensional quantile learning. Section \ref{sec:method} introduces our generative quantile method. The theoretical study in Section \ref{sec:thm} demonstrates the consistency of estimated the vector quantile and the posterior recovered from it. We  investigate the empirical performance of the proposed method in Section \ref{sec:nu}. Finally, we conclude the paper in Section \ref{sec:concluding}.


\section{Multivariate Quantile Learning}\label{sec:multi_q}
Quantile learning has a long history of literature in Statistics. A brief review of one-dimensional quantile learning is deferred to Section \ref{sec:additional_review} (Supplement). As meaningful ordering in $\R^d$ is not  obvious, so is extending the concept of quantiles, signs, and ranks from a univariate to a multivariate setting. 
For a comprehensive discussion of the various notions of a multivariate quantile, we refer to \cite{hallin2022measure}. Our work builds on one of the more recent optimal transport perspectives. 

\subsection{Optimal Transport for Quantile Learning}\label{sec:optim.transport}
The inherent ambiguity of the multivariate quantile arises from the fact that for a target distribution $P$, there exist multiple mappings $Q$ such that if $U$ follows some source distribution $F_U$ on the domain $\mathcal{U}$, then $Q(U)\sim P$;
any of these maps determines a transport from $F_U$ to $P$ \citep{koenker2017quantile}.
Recently, \cite{carlier2016vector} resolves the ambiguity of the mapping by adding an identifiability condition on $Q$ to be a gradient of a convex function, and call such $Q$  a {\em vector quantile} function.
This additional condition ensures the uniqueness of the mapping and can be viewed as a generalization of the monotonicity requirement for the univariate quantile function, as it leads to $[Q(u)-Q(u')]^\top (u-u')\geq 0$ for all $u,u'\in \mathcal{U}$.
The concept of vector quantiles was first introduced in the context of quantile regression   \citep{carlier2016vector} and   later extended by \cite{chernozhukov2017monge} and \cite{hallin2021distribution} to a general multivariate quantile through which data depth, ranks, and signs can be defined. Here, we follow developments in \cite{chernozhukov2017monge}.

Going forward in this paper, let the domain of $U$ be a $d$-dimensional unit Euclidean ball, i.e. $\mathcal{U} = S^d(1)$.
The source distribution $U \sim F_U$ is defined as follows: let $U = r\phi$, where $r\sim \unif([0,1])$, and $\phi\in\R^d$ follows uniform distribution on the $(d-1)$-dimensional unit sphere $\mathcal{S}^{d-1}(1)$, with $r$ and $\phi$ mutually independent.
 Consider a target distribution $P$ over $\R^d$.
The vector quantile mapping, denoted by $Q_P$, is then defined as a \emph{gradient of a convex function} which transforms $F_U$ onto the target distribution $P$.
It can be shown that such a map uniquely exists.\footnote{Theorem 2.1 in \cite{chernozhukov2017monge}, where the spherical uniform distribution $F_U$ is regarded as the source distribution $F$ therein. The condition is then $F$ is absolutely continuous with respect to the Lebesgue measure on $\R^d$.}
When $d=1$, $Q_P$ reduces to the standard quantile function. Furthermore, when $P$ has finite moments of order two, $Q_P$ is the Monge-Kantorovich (MK) optimal transport map from $F_U$ to $P$ that minimizes the expected quadratic cost, i.e.
\begin{equation}\label{eq:monge}
    Q_P = \argmin_{Q: Q \# F_U = P}\E_{U\sim F_U}\|Q(U)-U\|^2.
\end{equation}
Here, $Q \# F_U$ denotes the pushforward measure of $F_U$ under the mapping $Q$, and $\|\cdot\|$ denotes vector 2-norm.
We denote by $\psi$ the convex potential function that defines $Q_P$ through $Q_P=\nabla \psi$.
By the Kantorovich duality for the problem \eqref{eq:monge}, the potential function $\psi$ (and consequently $Q_P$) can be expressed through  
\begin{equation}\label{eq:dual}
   \psi= \arginf_{\varphi: \mathcal{U}\rightarrow \R\cup\{+\infty\}}\left( \int \varphi~ \d F_U + \int \varphi^* ~\d P \right),
\end{equation}
where $\varphi^*$ is a convex conjugate defined by $\varphi^*(\theta):=\sup_{u\in \mathcal{U}}[\theta^\top u - \varphi(u)]$.
Due to the celebrated Brenier's theorem \citep{brenier1991polar}, we know that the gradient of $\psi$ in \eqref{eq:dual} provides the optimal transport map, i.e., $\nabla \psi=Q_P$, and this gradient is referred to as the Brenier map.
In machine learning literature, apart from quantile perspectives, the convex potential $\psi$ in \eqref{eq:dual} has been learned by input convex neural networks (ICNNs) \citep{amos2017input} for generative modeling (See, Section \ref{sec:additional_ml_review}).

\begin{figure}
    \centering
    \includegraphics[width=0.32\linewidth]{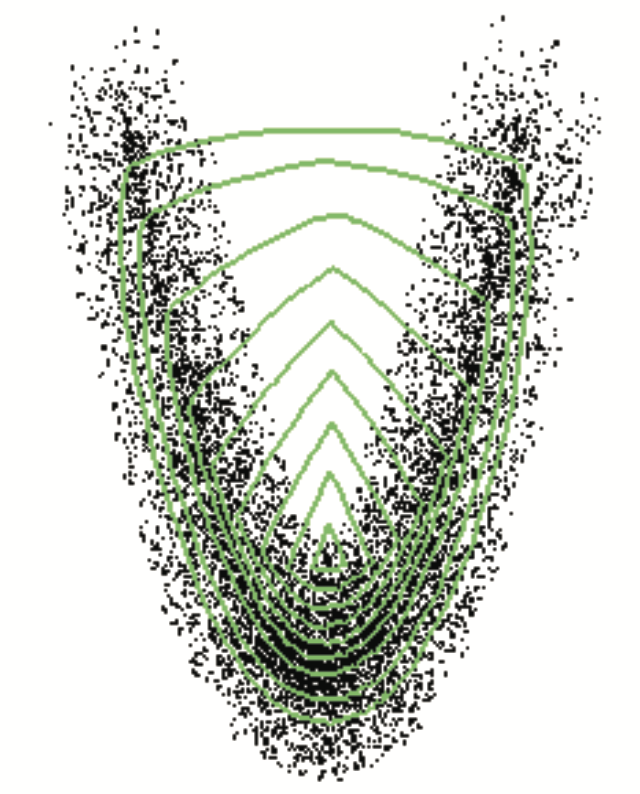}
    \includegraphics[width=0.35\linewidth]{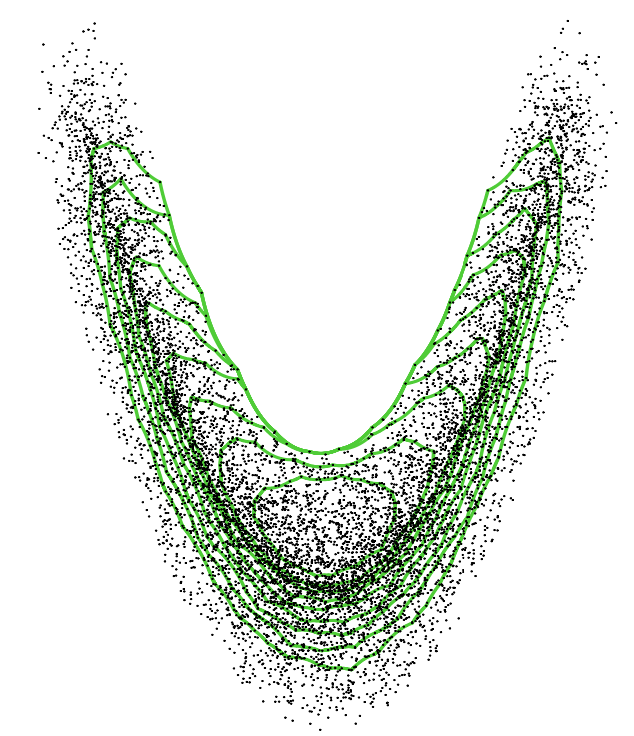}
    \caption{The data depth on a banana shaped distribution. Left: Halfspace depth \citep{tukey1975mathematics}. Right: Monge-Kantorovich depth \citep{chernozhukov2017monge}. The right one aligns more with our natural idea of credible set. Copied from \cite{chernozhukov2017monge}.}
    \label{fig:datadepth}
\end{figure}

\subsubsection{Conditional Vector Quantile}\label{sec:Conditional_details}
As the focus of our work is posterior sampling, our interest is more related to \emph{conditional} quantile learning by \cite{carlier2016vector}.
Note that our target function is now $\pi(\theta \mid X)$.
A conditional quantile map $Q_{\theta\mid X}(U)$ is defined as a map that satisfies the following properties: (1) for a fixed $X=x$, $Q_{\theta\mid X=x}(u)$ is a gradient of a convex function with respect to $u$, and (2) $Q_{\theta \mid X}(U)\sim \pi(\theta \mid X)$ for $U\mid X\sim F_U$ (given that $U$ is independent to $X$, it suffices to have $U\sim F_U$).
Similar to \eqref{eq:dual}, we can find a generative map by $Q_{\theta\mid X=x}(u) = \nabla_u \psi(u,x)$ (and its inverse map, denoted by $R_{u\mid X=x}(\theta) = \nabla_\theta \psi^*(\theta,x)$), where
\begin{align}\label{eq:cond_dual}
    \begin{split}
        \psi =& \arginf_{\varphi}\int\varphi(x,u)F_X(\d x)F_U(\d u)\\
        &+\int \varphi^*(x,\theta)F_{X,\theta}(\d x,\d \theta),
    \end{split}
\end{align}
and $\varphi^*(\theta,x) = \sup_{u\in \R^d} [\theta^\top u - \varphi(u,x)]$.
Here, $F_X$ is the marginal distribution of $X$, and $F_{X,\theta}$ is the joint distribution of $(X,\theta)$.
We can think $\nabla_u\varphi(x,u)$ as a Breinier map extended to a conditional version.

The primal problem of \eqref{eq:cond_dual} is a Monge problem,
\begin{equation}\label{eq:primal}
    \min_{U}\{\E\|\theta-U\|^2: ~U\sim F_U,~U \indep X\},
\end{equation} which can be seen as a conditional version of \eqref{eq:monge}. In \cite{carlier2017vector} and \cite{carlier2016vector}, a relaxed condition on the relationship between $X$ and $U$ and considered instead
\begin{equation}\label{eq:primal2}
    \min_{U}\{\E\|\theta-U\|^2: U\sim F_U,~\E[X\mid U] = \E[X]\}.
\end{equation} 
{
The additional mean-independence constraint $\E[X\mid U] = \E[X]$ replaces the conditional independence constraint $(U\indep X)$ in the Monge problem \eqref{eq:primal}.
The motivation is a simplified dual problem.
}

\subsubsection{Monge Kantorovich Data Depth}\label{sec:credible}

Data depth is a way of giving a weak ordering to a multidimensional data space. Denoting by $D_P(\theta)$ the depth of $\theta\in \R^d$ relative to $P$, we are given the depth ordering $\geq_{D_P}$ such that $\theta_1 \geq_{D_P}\theta_2$ if and only if $D_P(\theta_1)\geq D_P(\theta_2)$ for $\theta_1,\theta_2\in \R^d$.
The boundary of $\{\theta\in \R^d:D_P(\theta)\geq t\}$ gives us the contour of depth $t$.
The most well-known depth is Tukey's halfspace depth \cite{tukey1975mathematics}.
To avoid the situation when the depth region is defined outside of the support of the data, \cite{chernozhukov2017monge} instead apply Tukey's depth after pushing back $P$ to $F_U$.
See the representative illustration in Figure \ref{fig:datadepth}.
Here we explain this new depth notion by \cite{chernozhukov2017monge}, called MK-depth (for the definition of Tukey's halfspace depth, see \cite{chernozhukov2017monge}).
Denote by $\psi ^*$ the convex conjugate of $\psi$ defined in \eqref{eq:dual}.
\cite{chernozhukov2017monge} defines $R_P:=\nabla \psi^*$ as the vector rank function, which is the inverse of the quantile function\footnote{due to Brenier-McCann's theorem (Theorem 2.2 in \citep{chernozhukov2017monge})}, i.e., $R_P = Q_P^{-1}$.
Therefore, $R_P$ reverses $\theta\sim P$ back to $U\sim F_U$.
The MK depth is defined as the Tukey depth on the reversed distribution.
Interestingly, the MK-depth region with probability content $\tau\in(0,1)$ is identical to $Q_P(S^d(\tau))$.
Note that this MK-quantile region is controlled by a single $\tau\in(0,1)$ and has a nested structure $Q_P(S^d(\tau))\subset Q_P(S^d(\tau'))$ when $\tau\leq \tau'$.
Thus, we may see the MK-quantile region as another generalization of the one-dimensional quantile concept as a function of $\tau\in(0,1)$.

\vspace{-0.2cm}
\section{Generative Bayesian Computation}\label{sec:method}
{In this section, we extend the conditional vector quantile by \cite{carlier2016vector} to adopt summary statistics for Bayesian quantile learning. Then, we present our deep generative Bayes algorithm and implementation, along with the credible set computation.}

\subsection{Vector Quantile for Generative Bayes}\label{sec:vector.quantile.method}

Let $f: \mathcal{X}\rightarrow \R^q$ be any feature map such that $f(X)$ is a summary statistic separating $X$ and $\theta$, i.e. $X \indep \theta \mid f(X)$.
The dual problem of the relaxed \eqref{eq:primal2} has the form  
\begin{equation}\label{eq:dual3}
    \inf_{b,\varphi} \left(\int \varphi(u) F_U(\d u) + \int \tilde\varphi(f(x),\theta) F_{X,\theta}(\d x, \d\theta)\right),
\end{equation}
where $\tilde\varphi (f(x),\theta ) : = \max_{u \in \mathcal{U}} [u^\top \theta-\varphi(u)-b (u)^\top f(x)]$, with $b:\mathcal{U}\rightarrow\R^q$ and $\varphi: \mathcal{U}\rightarrow\R$ being continuous functions.
{ This dual problem, compared to \eqref{eq:cond_dual}, enables separation of $X$ and $U$.
In a similar argument to} \cite{carlier2017vector}, given $\E[f(X)] = 0$, for $U$ that solves the primal \eqref{eq:primal2} and $\varphi,b$ that solve \eqref{eq:dual3}, the following almost-sure relationship holds,
\[
\tilde\varphi (f(X),\theta)=U^\top \theta-\varphi(U)-b(U)^\top f(X).
\]
Thus the potential function takes the form of $\psi(u,x) = \varphi(u) + b (u)^\top f(x)$.
If $b$ and $\varphi$ are smooth, then the partial subdifferential of $\psi(u,x)$ with respect to $u$ becomes degenerate and collapses to the partial gradient $\nabla_u \psi(u,x)$.

It is also worth noting that the introduction of arbitrary summary statistics $f(X)$ represents an improvement over the original method by \cite{carlier2017vector}, which used $X$ in place of $f(X)$.
The latter relies on the assumption that the potential function
\begin{equation}\label{eq:old.affinity.assumption}
    \psi(u,x) = \varphi(u) + b(u)^\top x,
\end{equation}
which is equivalent to the vector quantile $Q_{\theta\mid X}$ being affine to $X$.
This specific requirement is rarely satisfied in practice.
Our analysis, on the other hand, 
{
relies on the following assumption.
\begin{assumption}[True Quantile Affinity]\label{assumption:2.7}
    There exists some summary statistics $f(X)$ that satisfies $X \indep \theta \mid f(X)$ and
    \begin{equation}\label{eq:affinity.assumption}
        \psi(u,x) = \varphi(u) + b (u)^\top f(x),
    \end{equation}
    or equivalently, $Q_{\theta\mid X}$ being affine to $f(X)$.
\end{assumption}
It is a feasible extension of \eqref{eq:old.affinity.assumption}, because there exists multiple summary statistics, usually an uncountable class up to some transformations.
More importantly, Assumption \ref{assumption:2.7} plays a central role in our methodology, and should not be simply regarded as a mere technical generalization of \eqref{eq:old.affinity.assumption}.
The goal of learning the potential function $\psi(u,x)$ is split into two sub-tasks, learning the summary statistics $f(x)$ and fitting functional coefficients $\varphi(u)$ and $b(u)$ that are convex.
This opens the door for a broad class of theoretically-guaranteed quantile learning methods, and turns out to be the cornerstone of our proposed algorithm.

}

\subsection{Deep Generative Quantile Bayes}
{Consider the parameter space for $\theta$ as $\Theta\subset \R^{d}$ and the data $X$ consists of $n$ observations of $d_X$-dimensional vectors $X=[x_1,...,x_n]$, where $x_i\in \R^{d_X}$, and the domain of $X$ is denoted by $\mathcal{X}\subset \R^{d_X\times n}$.
We assume that it is possible to simulate $(\theta, X)$ from the prior $\pi(\theta)$ and the likelihood $L(X\mid\theta)$ (which could also be implicit, i.e. likelihood-free models). Define $\bm \theta = \{\theta_i\}_{i=1}^N, ~\bm X = \{X_i\}_{i=1}^N,$ and $\bm U = \{U_i\}_{i=1}^N$, where $U_i\sim F_U$.}

We apply the method of \cite{sun2022conditional} from the perspective of \cite{wang2022adversarial} and \cite{polson2023generative}. We freshly simulate the training set (mini-batch) for every iteration, which is a set of triples $\{(\theta_i,X_i,U_i)\}_{i=1}^N$.

To train the functions $\varphi$, $b$, and $f$, we optimize the following objective function
\begin{align}\label{eq:sun_objective}
    \begin{split}
        \mathcal{L}_1&(\varphi,b,f \mid \bm X,\bm \theta,\bm U) = \sum_{i=1}^N \Big(\varphi(U_i) + \\& \max_{j\in \{1,...,N\}} \big\{U_j^\top \theta_i - \varphi(U_j)-b(U_j)^\top f(X_i)\big\}\Big), 
    \end{split}
\end{align}
which was first designed by \cite{sun2022conditional} to optimize the dual problem \eqref{eq:dual3} in Section \ref{sec:Conditional_details}. {The algorithm is presented in Algorithm \ref{alg:brenier}.}
The use of the objective function as in \eqref{eq:sun_objective} encourages learning a function $f(X)$ that is linear to the true quantile $Q_{\theta\mid X}$, based on which the linear coefficients $\varphi(u)$ and $b(u)$ are accurately learned.
To ensure that $f(X)$ is a summary statistic, an option is to learn $f$ with DeepSet, the details of which are elaborated in Section \ref{sec:summary.learning}.
Meanwhile, $\varphi$ and $b$ are parametrized using ICNNs as in \cite{sun2022conditional} to ensure their convexity in $u$ {(3 hidden layers of width 512, and CELU activation)}.
{
Any summary statistic $f(X)$ learned by Algorithm \ref{alg:brenier} that is not mean-zero will result in $b$ and $\varphi$ diverging to infinity.} To maintain $\E[f(X)]=0$, we adopt batch normalization as in \cite{sun2022conditional}.

\begin{figure}[!t]
\centering
\begin{minipage}{\linewidth}
\begin{algorithm}[H]
\caption{Generative Quantile Posterior Sampler}\label{alg:brenier}
\centering
\begin{tabular}{l l}
\multicolumn{2}{l}{\textbf{Fixed Input}: number of iterations $T$, the simul-}\\
\multicolumn{2}{l}{ation model $\pi(\theta) \times L(X\mid \theta_i)$} \\
\multicolumn{2}{l}{\textbf{Learnable Input}: Learnable networks $\varphi,b$ and $f$} \\
\multicolumn{2}{c}{\cellcolor[gray]{0.9}{\textbf{Training}}} \\
\multicolumn{2}{l}{For $t=1,\ldots,T$} \\
\multicolumn{2}{l}{\qquad Simulate $\{(\theta_i,X_i)\}_{i=1}^N$ from $\pi(\theta) \times L(X\mid \theta_i)$} \\
\multicolumn{2}{l}{\qquad Sample $\{U_i\}_{i=1}^N$ and $U_i \sim F_U$} \\
\multicolumn{2}{l}{\qquad Compute $\mathcal{L}_1$ as in \eqref{eq:sun_objective}} \\
\multicolumn{2}{l}{\qquad Update $\varphi, b$ and $f$ via a stochastic optimizer} \\
\multicolumn{2}{c}{\cellcolor[gray]{0.9}{\textbf{Sampling $\tau$-credible set conditioned on $X=x$}}} \\
\multicolumn{2}{l}{Sample $U \sim F_U$} \\
\multicolumn{2}{l}{Compute $\psi(U,x) = \varphi(U)+b(U)^\top f(x)$} \\
\multicolumn{2}{l}{Return posterior sample $\nabla_U\psi(U,x)$}
\end{tabular}
\end{algorithm}
\end{minipage}
\end{figure}

In our implementation, we use Adam optimizer with default hyperparameter settings, and with a learning rate 0.01. For every epoch (every 100 iterations), we reduce the learning rate multiplicative by 0.99. 

\subsection{Automatic Learning for Summary Statistics} \label{sec:summary.learning}
To learn the feature map $f:\mathcal{X}\rightarrow \R^{q}$ from data to $q$ summary statistics, we consider DeepSet $h_1: \mathcal{X}\rightarrow \R^{q_1}$ and LSTM $h_2:\mathcal{X}\rightarrow \R^{q_2}$. Then, our feature map is defined as $f(X) = [h_1(X),h_2(X)]$, where $q=q_1+q_2$. Here $h_1$ is to represent order invariant summary statistics, while for $h_2$, the order of observation matters. For example, when we know that the data is i.i.d., we set $q_2=0$; Otherwise, we can set $q_2>0$ to reflect the possible existence of dependence among observations, e.g., sample autocorrelations. 

Our design of $f(\cdot)$ through DeepSet and LSTM has a potential for scaling up other deep learning based Bayesian methods  including \cite{wang2022adversarial} and \cite{kim2023deep}. When the standard fully connected neural network is used, the network size (number of network parameters to optimize) scales up with the input dimension, which is not realistic for a large $n$. The deep set design was also adopted for neural estimators, e.g., for extreme value analysis \citep{sainsbury2024likelihood} and spacial data analysis \citep{richards2023neural}.

{
\subsection{Credible Set Computation}
}
Generative quantile posterior learning enables direct sampling from multivariate posterior credible sets. Existing samplers like MCMC and ABC methods compute these sets using posterior draws based on a chosen metric, after committing to certain geometry of the set, e.g. ellipsoid.
{
On the contrary, our approach does not impose any specific restriction on geometric structure, and is able to automatically learn the shape of the credible sets.
}
As proposed by \cite{chernozhukov2017monge} and \cite{hallin2021distribution}, the vector quantiles define data depth, from which we can derive the depth region (the deepest set) and quantile contours. Depth region of probability $\tau$ then can be used as a credible set of probability $\tau$.
{
Thanks to these desirable properties of the MK depth, we shall see in Section \ref{sec:thm} that our approach is asymptotically valid, in the sense that the credible sets converge to the oracle set derived from the true underlying posterior.
}

\section{Theoretical Studies}\label{sec:thm}
The simulated training data $\{\theta_i, X_i, U_i\}_{i=1}^N$ consists of  i.i.d. observations.
However, there is no requirement on the dependence structure within $X_i = (X_{i1},\cdots,X_{in})^\top$.

Recall that our main goal is simulating from an approximated mapping to the posterior distribution of $\theta$ given the observed data vector $X$.
Let $U\sim F_U$ be a random vector that follows the source distribution (as given in Section \ref{sec:optim.transport}) on a $d$-dimensional unit Euclidean ball $\mathcal U=S^d(1)$.
We aim to find a mapping $H: \mathcal{U} \times \mathcal{X} \rightarrow \Theta$  such that $H(U,X) \sim \pi(\theta \mid X)$.
This pursuit is warranted by the Noise Outsourcing Lemma (see Theorem 5.10 in \cite{kallenberg2002foundations}).
In the presence of a summary statistic $f(X)$ such that $X$ and $\theta$ are conditionally independent given $f(X)$, this lemma  (refined by \cite{bloem2020probabilistic}) reads as follows:
\begin{lemma}[Noise Outsourcing Lemma]\label{lm:noise.outsourcing}
    Let $\mathcal{S}$ be a standard Borel space and $f:\mathcal{X}\rightarrow\mathcal{S}$ a measurable map.
    Then $X\indep \theta \mid f(X)$ if and only if there is a measurable function $H: \mathcal{U} \times \mathcal{S} \rightarrow \Theta$ such that
    \begin{equation}\label{eq:noise.outsourcing}
        (X,\theta) \stackrel{a.s.}{=} (X, H(U, f(X))),
    \end{equation}
    where $U \sim F_U$ and $U\indep X$.
    In particular, $\theta = H(U, f(X))$ has  a distribution $\pi(\theta\mid X)$.
\end{lemma}
A requirement on the source distribution in Lemma \ref{lm:noise.outsourcing} is that the support of $U$ must be sufficient to allow $H(U,f(X))$ to attain all values in the support of $\Theta$ given the summary statistics $f(X)$.
This requirement is satisfied since our source distribution $F_U$ is continuous and has a positive density anywhere in the unit ball $S^d(1)$.
Any mapping that satisfies \eqref{eq:noise.outsourcing} can be regarded as a conditional vector quantile function and a conditional generator for $\pi(\theta\mid X)$.
Lemma \ref{lm:noise.outsourcing} guarantees that such a mapping exists.

In our methodology, the conditional vector quantile $Q_{\theta\mid X=x}$ stems from a potential function $\psi(u,x)$ that is convex in $u$.
Likewise, the inverse of such a vector quantile, i.e. conditional vector rank denoted by $R_{U\mid X=x}$ relies on $\psi^*(\theta,x)$, the conjugate of $\psi(u,x)$.
Specifically, for each $u\in\mathcal{U}$ and $\theta\in\Theta$, 
\begin{align*}
    \psi(u,x) &= \sup_{\theta\in\Theta}\left[\theta^\top u - \psi^*(\theta,x)\right],\\
    \psi^*(\theta,x) &= \sup_{u \in \mathcal{U}}\left[\theta^\top u - \psi(u,x)\right].
\end{align*}
The following condition is imposed on the pair of potentials for a rigorous definition of $Q_{\theta\mid X=x}$ and $R_{U\mid X=x}$.
{
It ensures that the vector quantile and vector rank are well-defined and are the inverse of each other.
}
\begin{condition}[Similar to Condition (C) in \cite{chernozhukov2017monge}]\label{cond:C}
    There exist open, non-empty subsets $\mathcal{U}_0 \subset \mathcal{U}$ and $\Theta_0\subset\Theta$ such that
    \begin{enumerate}
        \item $\psi$ and $\psi^*$ possess gradients $\nabla_u \psi(u,x)$ for all $u\in\mathcal{U}_0$, and $\nabla_\theta \psi^*(\theta,x)$ for all $\theta\in\Theta_0$, respectively,
        \item the restrictions $\nabla\psi |_{\mathcal{U}_0}: \mathcal{U}_0 \rightarrow \Theta_0$ and $\nabla \psi^* |_{\Theta_0}: \Theta_0 \rightarrow \mathcal{U}_0$ are homeomorphisms,
        \item $\nabla\psi |_{\mathcal{U}_0} = (\nabla \psi^* |_{\Theta_0})^{-1}$.
    \end{enumerate}
\end{condition}
Under our settings, it is natural to require $\mathcal{U}_0 = \text{int }\mathcal{U}$, i.e. the interior of the unit ball $S^d(1)$, and $\Theta_0 = \text{int }\Theta$.
Vector quantiles and vector ranks are thus defined as follows:
\begin{align*}
    &Q_{\theta\mid X=x}(u) = \arg\sup_{\theta\in \Theta}\left[\theta^\top u - \psi^*(\theta,x)\right],\quad u\in \mathcal{U}_0;\\
    &R_{U\mid X=x}(\theta) = \arg\sup_{u\in \mathcal{U}}\left[\theta^\top u - \psi(u,x)\right],\quad \theta\in \Theta_0.
\end{align*}
By the envelope theorem and Rademacher's theorem \citep{rademacher1919partielle},
\begin{align}
    Q_{\theta\mid X=x}(u) &= \nabla_u \psi(u,x)\quad \text{a.e. on }\mathcal{U}_0,\\
    R_{U\mid X=x}(\theta) &= \nabla_\theta \psi^*(\theta,x)\quad \text{a.e. on }\Theta_0.
\end{align}
Since the distribution of $U$ is absolutely continuous with respect to the Lebesgue measure on $\R^d$, the results in \cite{brenier1991polar} and \cite{mccann1995existence} show the existence of function $\psi: \mathcal{U} \times \mathcal{X}\rightarrow \R \cup \{+\infty\}$ that is convex in $u$, and that its gradient $\nabla_u\psi$ is unique almost everywhere (see Theorem 2.1 in \cite{chernozhukov2017monge}).
Similarly, the existence of $\psi^*(\theta,x)$ and the uniqueness of $\nabla_\theta \psi^*$ are also guaranteed.

Recall the affinity assumption \eqref{eq:affinity.assumption}.
Our goal is to estimate the coefficients $\varphi, b$ and the summary statistic $f$ using the generated data flow $\{(X_i, \theta_i, U_i)\}_{i=1}^N$. Specifically,
\[
\hat{\psi}_N(u,x) = \hat{\varphi}_N(u) + \hat{b}_N(u)^\top \hat{f}_N(x),
\]
where $\hat{\varphi}_N$, $\hat{b}_N$, and $\hat{f}_N$ are learned from a feed-forward neural network.
The convex conjugate of $\hat{\psi}_N(u,x)$ is defined as
\[
\hat \psi_N^*(\theta,x) = \sup_{u\in \mathcal{U}} \{\theta^\top u - \hat \psi_N(u,x)\}.
\]
Note that the subscripts $N$ corresponds to the estimates stemming from the generated datesets.
Our estimate of vector quantile and vector rank functions is defined as follows:
\begin{align*}
    &\hat{Q}^N_{\theta\mid X=x}(u) \in \arg\sup_{\theta\in \Theta}\left[\theta^\top u - \hat{\psi}_N^*(\theta,x)\right],\quad u\in \mathcal{U};\\
    &\hat{R}^N_{U\mid X=x}(\theta) \in \arg\sup_{u\in \mathcal{U}}\left[\theta^\top u - \hat{\psi}_N(u,x)\right],\quad \theta\in \Theta.
\end{align*}
Thus we have
\begin{align*}
    \hat{Q}_{\theta\mid X=x}^N(u) &= \partial_u \hat{\psi}_N(u,x)\quad \text{for } u\in \mathcal{U},\\
    \hat{R}^N_{U\mid X=x}(\theta) &= \partial_\theta \hat{\psi}_N^*(\theta,x)\quad \text{for } \theta\in \Theta,
\end{align*}
where $\partial$ denotes the subdifferential of a convex function.

To quantify the approximation error of $\hat{\psi}_N(u,x)$, we apply the generic result in \cite{farrell2021deep} regarding the approximation error of functionals using feed-forward neural networks.
The following condition is required.
\begin{condition}\label{cond:farrell.1}
    There exists an absolute constant $M>0$, such that
    \begin{enumerate}
        \item the true underlying functions satisfy $\| \varphi \|_\infty \leq M$, $\| b_k \|_\infty \leq M$, $\| f_k \|_\infty \leq M$ for all $1\leq k\leq q$, and the range $\Theta \subset [-M,M]^d$.
        \item The estimates of the functions from the feed-forward neural networks are uniformly bounded, i.e. for any $N\in\N$, $\|\hat{\varphi}_N\|_\infty \leq 2M$, $\|\hat{b}_{N,k}\|_\infty \leq 2M$, $\|\hat{f}_{N,k}\|_\infty \leq 2M$ for all $1\leq k\leq q$.
        \item The estimates of the functions from the feed-forward neural networks are uniformly equicontinuous on the compact domain $\mathcal{U}$ or $\mathcal{X}$, i.e. for any $\epsilon>0$, there exists $\delta>0$ such that for any $N>0$, any $u_1,u_2\in\mathcal{U}$ such that $\|u_1 - u_2\| <\delta$ implies $\|\hat{\varphi}_N(u_1) - \hat{\varphi}_N(u_2)\| < \epsilon$ and $\|\hat{b}_{N,k}(u_1) - \hat{b}_{N,k}(u_2)\| < \epsilon$, and any $x_1, x_2 \in\mathcal{X}$ such that $\|x_1 - x_2\| < \delta$ implies $\|\hat{f}_{N,k}(x_1) - \hat{f}_{N,k}(x_2)\| < \epsilon$.
    \end{enumerate}
    
\end{condition}
{
These boundedness and compactness assumptions are fairly standard in nonparametrics; see \cite{farrell2021deep}.
The choice of $M$ may be arbitrarily large, and no properties of $M$ is required apart from being finite.}
We also impose the following condition on the settings of feed-forward networks for the learning of the functions.
\begin{condition}\label{cond:farrell.3}
    Let the true underlying functions $\varphi$, $b_k$, and $f_k$ for $1\leq k\leq q$ lie in classes $\mathcal{F}^\varphi$, $\mathcal{F}^b$, and $\mathcal{F}^f$, respectively.
    For the feed-forward network classes $\mathcal{F}_{\text{DNN}}^\varphi$, $\mathcal{F}_{\text{DNN}}^b$, and $\mathcal{F}_{\text{DNN}}^f$, let the approximation error be
    \begin{align*}
        \epsilon_{\text{DNN}}^\varphi &:= \sup_{\varphi\in\mathcal{F}^\varphi} \inf_{\hat{\varphi}_N\in\mathcal{F}_{\text{DNN}}^\varphi, \|\varphi\|\leq 2M} \|\hat{\varphi}_N - \varphi\|_\infty;\\
        \epsilon_{\text{DNN}}^b &:= \max_{1\leq k\leq q}\sup_{b\in\mathcal{F}^b} \inf_{\hat{b}_{N,k}\in\mathcal{F}_{\text{DNN}}^b, \|b_k\|\leq 2M} \|\hat{b}_{N,k} - b_k\|_\infty;\\
        \epsilon_{\text{DNN}}^f &:= \max_{1\leq k\leq q}\sup_{f\in\mathcal{F}^f} \inf_{\hat{f}_{N,k}\in\mathcal{F}_{\text{DNN}}^f, \|f_k\|\leq 2M} \|\hat{f}_{N,k} - f_k\|_\infty.
    \end{align*}
    The network classes are selected such that as $N\rightarrow\infty$, all these quantities converge to zero.
\end{condition}
{
This condition assumes that the underlying functions are learned by the feed-forward neural network accurately enough.
For example, if the neural network structure is a multi-layer perceptron, a structure known as a good approximator of smooth function, the approximation errors $\epsilon_{DNN}^\varphi$, $\epsilon_{DNN}^b$, and $\epsilon_{DNN}^f$ could have a finite-sample upper bound, given that the true underlying functions belong to certain classes like Sobolev ball; see, e.g. \citep{gine2021mathematical}.
}
The following lemma shows a convergence result for the potential function $\hat{\psi}_N$ and its conjugate, $\psi_N^*$.
\begin{lemma}\label{lm:psi.consistency}
    { Under Assumption \ref{assumption:2.7},}
    suppose that Conditions \ref{cond:farrell.1} and \ref{cond:farrell.3} hold.
    Then almost surely for the generated data flow $\{\theta_i, X_i, U_i\}_{i=1}^N$, the estimated potential function $\hat{\psi}_N(u,x)$ and its conjugate $\hat{\psi}_N^*(\theta,x)$ satisfy the follwing. 
    \begin{enumerate}
        \item $\hat{\psi}_N(u,x)$ converges uniformly to $\psi(u,x)$ over $\mathcal{U}\times\mathcal{X}$, i.e.
        \begin{equation}
            \lim_{N\rightarrow\infty} \sup_{(u,x)\in \mathcal{U}\times\mathcal{X}} |\hat \psi_N(u,x) - \psi(u,x)| = 0;
        \end{equation}
        \item For any compact set $K' \subset \Theta$, $\hat \psi_N^*(\theta,x)$ converges uniformly to $\psi^*(\theta,x)$ over $K'\times\mathcal{X}$, i.e.
        \begin{equation}
            \lim_{N\rightarrow\infty} \sup_{(\theta,x)\in K'\times\mathcal{X}} |\hat \psi_N^*(\theta,x) - \psi^*(\theta,x)| = 0;
        \end{equation}
    \end{enumerate}
    
\end{lemma}

We now present the main theorem of this section, which provides an asymptotic consistency guarantee for the estimated vector quantile $\hat Q^N_{\theta\mid X}$, as well as the recovered posterior $\hat{\pi}^N(\theta\mid X)$.
In a slight abuse of notation, denote the 2-Wasserstein distance between the probability measures of the recovered posterior and the true posterior by $W_2(\hat \pi^N(\theta \mid X), \pi(\theta\mid X))$.
\begin{theorem}\label{thm:main}
    { Under Assumption \ref{assumption:2.7},}
    suppose that Conditions \ref{cond:C}, \ref{cond:farrell.1} and \ref{cond:farrell.3} hold.
    Then almost surely for the generated data flow $\{\theta_i, X_i, U_i\}_{i=1}^N$, the following consistency results hold.
    \begin{enumerate}
        \item For any closed subset $K_U \subset \mathcal{U}_0$ and $K_X \subset \mathcal{X}$ as $N\rightarrow\infty$,
        \begin{equation}\label{eq:main.thm.1}
            \sup_{u\in K_U, x\in K_X} \| \hat{Q}^N_{\theta\mid X=x}(u) - Q_{\theta\mid X=x}(u) \| \rightarrow 0.
        \end{equation}

        \item Furthermore, for any closed subset $K_X \subset \mathcal{X}$,
        \begin{equation}\label{eq:main.thm.2}
            \sup_{x\in K_X} W_2(\hat{\pi}^N(\theta \mid X=x), \pi(\theta\mid X=x)) \rightarrow 0.
        \end{equation}
    \end{enumerate}
\end{theorem}
Note that when $\mathcal{X}$ itself is closed, the supremum over $x\in K_X$ can be written as over $x\in \mathcal{X}$.

A key corollary of Theorem \ref{thm:main} is the consistency of Bayesian credible sets.
Given any $\tau\in(0,1)$, denote the $d$-dimensional ball with radius $\tau$ as $S^d(\tau)$.
Recall Section \ref{sec:credible} that the Bayesian credible sets are constructed from the estimated Monge-Kantorovich vector quantile as follows,
\begin{align}\label{eq:credible.set}
    \hat C^N_\tau (\theta \mid X=x) &:= \hat Q^N_{\theta\mid X=x}(S^d(\tau)),
\end{align}
while under the true posterior, the oracle set is defined as
\begin{align}\label{eq:oracle.set}
    C_\tau (\theta \mid X=x) &:= Q_{\theta\mid X=x}(S^d(\tau)).
\end{align}
The following corollary verifies that the credible set $\hat C^N_\tau (\theta \mid X=x)$ converges to $C_\tau (\theta \mid X=x)$, in terms of Hausdoff distance,
\[
d_H(A,B) = \max\left\{\sup_{a\in A}\inf_{b\in B}\|a-b\|, \sup_{b\in B}\inf_{a\in A}\|a-b\|\right\}.
\]
\begin{corollary}\label{corr:credible.set}
    Suppose that the conditions of Theorem \ref{thm:main} hold.
    Then the generated data flow $\{X_i, \theta_i, U_i\}_{i=1}^N$ almost surely yields an estimated quantile $\hat Q^N_{\theta\mid X}$ such that for any $\tau\in(0,1)$, the Bayesian credible set in \eqref{eq:credible.set} satisfies
    \begin{equation}
        \sup_{x\in K_X}d_H(\hat C^N_\tau (\theta \mid X=x), C_\tau (\theta \mid X=x)) \rightarrow 0
    \end{equation}
    as $N\rightarrow \infty$, for any closed subset $K_X\subset\mathcal{X}$.
\end{corollary}

An important remark is related to support shrinkage.
As the number of observations $n$ increases, contraction occurs on the true underlying posterior, leading to a contraction of the oracle sets.
If all the assumptions we have made so far are satisfied,
{
then Corollary \ref{corr:credible.set} implies that the Bayesian credible sets should shrink similarly as the oracle sets do.
}
The violation of these assumptions may be the reason why support shrinkage was not observed in many previous methods,
{
a sign of the credible sets not converging to the oracle sets.
}

{
The theoretical analysis in this section is not particularly tailored to our method in Section \ref{sec:method}, but rather more generic.
As a supplement to the discussion in Section \ref{sec:vector.quantile.method}, we would like to emphasize that Assumption \ref{assumption:2.7} is motivated by Lemma \ref{lm:noise.outsourcing}, the Noise Outsourcing Lemma.
On a broader scale, the wide range of quantile learning methods enabled by this assumption can be theoretically consistent, as long as they learn the summary statistics $f(X)$ and convex functions $\varphi(u)$ and $b(u)$ simultaneously, and the technical conditions are satisfied.
}

\vspace{-0.2cm}
\section{Numerical Studies}\label{sec:nu}
\subsection{Gaussian Conjugate Simulation}

\begin{figure}[t]
    \centering
    \includegraphics[width=\linewidth]{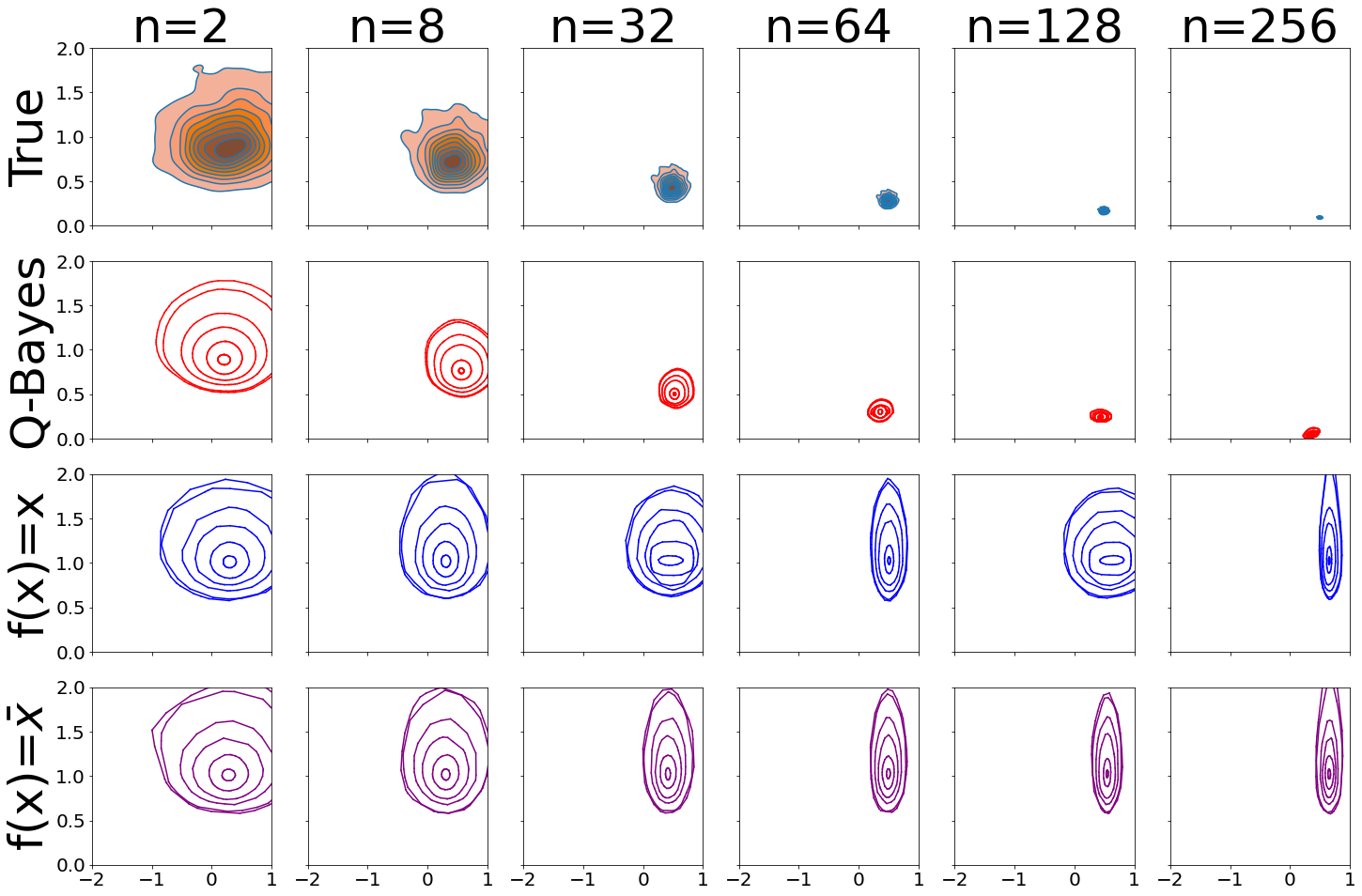}
    \caption{The credible set for the Gaussian example with increasing $n$ for a very small $x$ ($x=0.5$).}
    \label{fig:good}
\end{figure}

Consider a normal-inverse gamma model, where $X\mid \mu, \sigma^2\sim N(\mu,\sigma^2)$ with priors $\nu_0\sigma^2_0/\sigma^2\sim \chi^2(\nu_0)$ and $\mu\mid \sigma^2\sim N(\mu_0, \sigma^2/\kappa)$. { We choose this model, as it can be seen as one of the most standard posterior sampling examples, for which we know the true posterior distribution.} We set $\mu_0 = 0$, $\sigma_0=1$, $\kappa=2$, and $\nu_0=25$. Here, we increased $n$ using a DeepSet feature extractor for a few chosen $X=x$ values. With the DeepSet feature network (order-invariant network design), we can see that our method scales with increasing $n$ values. We highlight that in Figure \ref{fig:good} {(the second row)}, the support shrinkage (contraction of the estimated posterior contour sets) is clearly observed with an increasing $n$  for $x$ relatively near the origin. {In this figure, we can also see the effect of using DeepSet in comparison to having no feature extractor ($f(x) = x$) or insufficient statistics ($f(x) = \bar{x})$.} In Section \ref{sec:gauss} in Appendix, we provide more details of the experiment, comparison with B-GAN \citep{wang2022adversarial} and the Autoregressive method, the effect of the network choice  for $n=2$.

\vspace{-0.2cm}
\subsection{Brock Hommes Model} 

\begin{figure}
    \centering
    \includegraphics[width=0.8\linewidth]{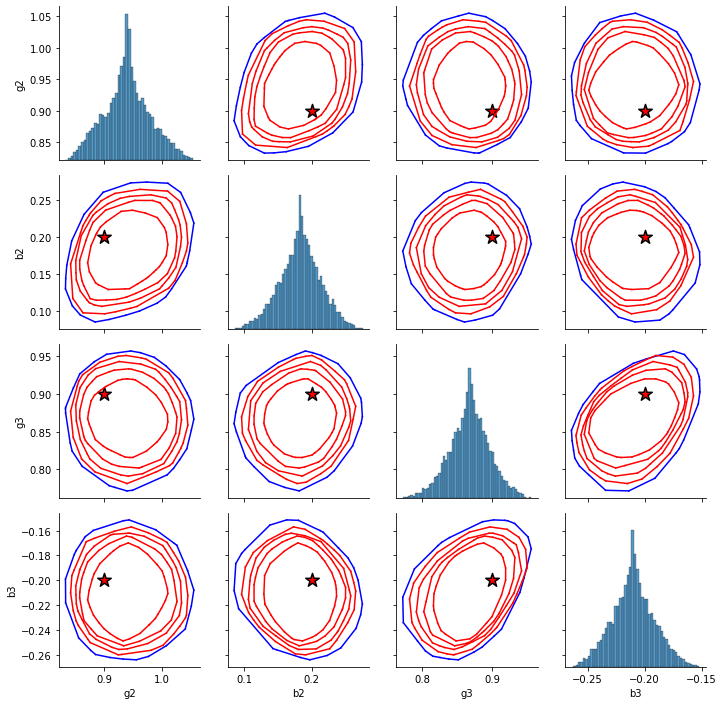}
    \caption{The Brienier map results on Brock Hommes Model, with convex hull corresponding to $\tau\in [0.5,0.6, 0.7, 0.8, 0.9]$ (red) and $\tau=1$ (blue). Red stars: the true parameters.}
    \label{fig:brock}
\end{figure}

\cite{brock1998heterogeneous} developed an agent-based model to simulate asset trading on an artificial stock market, capturing interactions among heterogeneous traders who follow various trading strategies. {  The Brock and Hommes model is one of the most foundational economic agent-based models and has been widely used due to its simplicity, while effectively incorporating heterogeneous agents. }Recently, \cite{platt2020comparison} applied this model to evaluate economic agent-based model calibrations. The model is
\begin{align*}
    x_{t+1} &= \frac{1}{R} \sum_{h=1}^{H} n_{h, t+1} (g_h x_t + b_h) + \epsilon_{t+1}, \\
    n_{h, t+1} &= \frac{\exp(\beta U_{h, t})}{\sum_{h=1}^{H} \exp(\beta U_{h, t})}, \\
    U_{h, t} &= (x_t - R x_{t-1}) (g_h x_{t-2} + b_h - R x_{t-1}),
\end{align*}
where $\epsilon_t \sim \mathcal{N}(0, \sigma^2)$, and $R, \beta, \sigma$ are parameters. Following \cite{platt2020comparison}, we set $\beta = 120$, $H = 4$, $R = 1.01$, $\sigma = 0.04$, $g_1 = b_1 = b_4 = 0$, and $g_4 = 1.01$. Additionally, as in \cite{dyer2024black}, we estimate the posterior $p(\theta \mid \mathbf{y})$, where $\theta = (g_2, b_2, g_3, b_3)$, $\mathbf{y} := (y_1, y_2, \dots, y_T) \sim p(\mathbf{x} \mid \theta^*)$ represents the pseudo-observation, with $T = 100$, and $\theta^* = (g_2^
*, b_2^*, g_3^*, b_3^*) = (0.9, 0.2, 0.9, -0.2)$ as the parameters used to generate $\mathbf{y}$.
The priors are specified as $g_2, b_2, g_3 \sim \mathcal{U}(0, 1)$, while $b_3 \sim \mathcal{U}(-1, 0)$. For further details on the model, including its interpretation, we refer readers to \cite{platt2020comparison}. The results of our method on this data are presented in Figure \ref{fig:brock}, which visualizes the posterior contour sets estimated from 10,000 generated samples. It is observed that these contour sets do not overlap. However, as this is an approximation, some overlap could occur. The parameters are contained within the 50\% credible set across all dimensions (the innermost contour).

{ In Section \ref{sec:table_detail} in Appendix, we also present a comparison with other methods such as the standard ABC (rejection ABC) and sequential Monte Carlo ABC (SMC-ABC, \cite{sisson2007sequential}), which show the competency of our method in terms of the quality of samples and computational time.}

\section{Concluding Remarks}\label{sec:concluding}
This paper develops an implicit sampler from a posterior distribution over multivariate parameters through quantile learning.  This methods scales with the number of (potentially dependent) observations and exhibits {\em support shrinkage}, i.e. shrinkage of the posterior approximation with $n$. In addition, we provide a   tool for estimating  contours of the posterior approximation  (including posterior credible sets) without imposing a strong geometric structure. Such a structure would be required for multivariate credible sets with more traditional sampling methods such as MCMC or ABC. { It is important to note, however, that our method fundamentally differs from both ABC and MCMC approaches.
Once our posterior generator is trained, it can be applied to any realization of datasets without the need for retraining.
In contrast, both MCMC and ABC must be re-run for each new dataset, which significantly increases their computational burden in practical applications.
This reusability makes our approach particularly advantageous for scenarios where multiple datasets need to be analyzed efficiently.}

{Our work focuses on models with intractable likelihoods and low-dimensional continuous parameter spaces. It would be interesting to extend this work to discrete parameter spaces. The training data is not tailored to the observed data $X_0$. This could be improved by constructing a dataset with samples more similar to $X_0$, such as using the approach in \cite{o2024tree}, which assigns importance weights to training observations. These weights would then be incorporated into the learning criterion.}
{
\acknowledgments {The authors sincerely acknowledge Nick Polson and Sehwan
Kim for their valuable discussions and insights. The authors gratefully acknowledge the support from the National Science Foundation (Grant number DMS 1944740).}}
\bibliographystyle{chicago}
\typeout{}
\bibliography{ref}

\appendix
\section*{Appendix}
\section{Comparison to Adversarial Bayesian Simulation} \label{sec:diff}Our approach is related to adversarial Bayesian simulation, or Bayesian GAN (B-GAN),  by \cite{wang2022adversarial}. Like B-GAN, our method bypasses the need for posterior sampling by learning a generative joint distribution through forward simulation. While B-GAN employs the 1-Wasserstein GAN framework, our work extends this to the 2-Wasserstein GAN framework, paralleling the relationship between W-GAN \citep{arjovsky2017wasserstein} and 2-Wasserstein GAN \citep{taghvaei20192}. In the original W-GAN (using the 1-W distance), the critic is trained to compute the distance between the current generating distribution and the data distribution. As the generator is trained to minimize this distance, the critic is iteratively optimized uner an adversarial learning process. In practice, the critic may not be fully optimized due to the iterative adversarial nature \citep{stanczuk2021wasserstein}. \cite{taghvaei20192} instead adopt the 2-Wasserstein distance, for which computing the distance only one time provides the generator, which is the derivative of the potential function (the Brenier map). Therefore, the need for iterative distance reduction is eliminated. If the potential function function is designed with fully input-convex networks, no additional regularization to maintain convexity is needed. Conversely, for the original WGAN, ensuring that the critic remains Lipschitz continuous with a constant of 1 is a challenging problem, which has motivated many subsequent works \citep{gulrajani2017improved, petzka2017regularization}. Another advantage of extending from the 1-Wasserstein to the 2-Wasserstein framework is the ability to separately design a feature map $f$ as described in \eqref{eq:sun_objective}, which can be treated as the condition and summary statistics. For the 1-Wasserstein framework, an interesting direction for future research would be to develop an algorithm where both the critic and the generator are conditioned on a learnable feature.

\section{Additional Literature Review}
\subsection{One-Dimensional Quantile Learning}\label{sec:additional_review} 
For the one-dimensional case, many generative approaches walk around the pinball loss defined by $\Lambda_\tau(q,z) = (\tau-\mathcal{I}_{(z<q)})(z-q).$ Denote the cdf by $F_z(\tau) = P(Z<\tau).$ The key property of this loss is that $$\argmin_{q}\E_{z\sim F_z}[\Lambda_\tau (q,z)] = F^{-1}_Z(\tau).$$ The generative models are trained by minimizing the continuous ranked probability score (CRPS, \cite{matheson1976scoring}), which is the pinball loss integrated over all quantile levels $\tau\in[0,1]$ defined by 
$${\rm CRPS}(F^{-1},z) = \int_0^1 2\Lambda_\tau(F^{-1}(\tau), z)\d \tau.$$ The model trained by it is sometimes called ``implicit'' quantile function, because $\tau$ is not fixed, but also fed into the model \citep{dabney2018implicit}. CRPS is a proper scoring rule \citep{gneiting2007strictly}, i.e., $\int g(z) {\rm CRPS}(G^{-1}, z) \d z \leq \int g(z) {\rm CRPS}(F^{-1}, z) \d z,$ for any distributions $F$ and $G$ (with $g$ being the density of $G$). 

In the context of quantile regression, where the conditional $F_{y\mid X}^{-1}(\tau)$ is pursued, \cite{wang2024generative} noticed that the trained model $\hat{F}$ may learn to output $\hat{F}_{y\mid X}^{-1}(\tau)=y$ for all $\tau\in[0,1]$. We think this happens because for each $X$ observation, only one $y$ might be observed, allowing the function to interpolate the training data. Therefore, \cite{wang2024generative} propose to use a penalty preventing this quantile collapse to a single  point. In settings learning with simulated data, the same problem may not occur, not needing the same preventive penalty.

\subsection{Optimal Transport in GAN}\label{sec:additional_ml_review}
While statistical literature provides theoretically rigorous quantile notions and in-depth understandings of them, in machine learning literature, much algorithmic effort has been made to promote the practicality and scalability of quantile learning. In machine learning, multivariate quantile learning has been approached from three different angles: through optimal transport \citep{sun2022conditional,makkuva2020optimal}, copula modeling \citep{zeng2022neural,wen2019deep}, and auto-regressive modeling \citep{koenker2006quantile, ostrovski2018autoregressive}. Here, we focus on the optimal transport perspective.

In developing 2-Wasserstein GAN, \cite{taghvaei20192} paid attention to \cite{villani2021topics} Theorem 2.9 (see, \cite{makkuva2020optimal}), 
\begin{equation}\label{eq:wass2}
    W_2^2(P,Q) = C_{P,Q} -\inf_{f: ~convex} \{\E_P[f(X)]+\E_Q[f^*(X)]\},
\end{equation}
where $f^*(y) = \sup_x\langle x,y\rangle - f(x)$ is the convex conjugate of $f(\cdot).$ By Brenier's theorem, the optimal transport w.r.t. $W_2^2(P,Q)$ is $\nabla f^*(y)$. The convex potential $f$ has been modeled by using input convex neural networks (ICNNs) \citep{amos2017input} as a one dimensional potential function, and the corresponding optimal transport has been used either as the generator or refiner of the generator (similar idea was used independently by \cite{tanaka2019discriminator}). \cite{makkuva2020optimal} stabilized the computation of \eqref{eq:wass2}, replacing the role of $f^*$ by another convex function $g$, by $ W_2^2(P,Q) =  C_{P,Q} -\sup_{f}\inf_{g} \{\E_P[f(X)]+\E_Q[\langle Y, \nabla_g(Y)\rangle - f(\nabla_g (Y))],$ where $f$ and $g$ are convex. \cite{huang2020convex} further use the gradient of the convex potential $\nabla f$ as a Flow model, by developing the inversion algorithm of $\nabla f$ and computations needed for the hessian computation for the likelihood calculation. When $P$ is set a uniform $V\sim U(0,1)^d$ or other valid distributions of $V$ in \cite{carlier2016vector}, these generative approaches can be seen as multi-dimensional quantile modelling (not conditional).

\section{Proof of Theorems}

This section gives the proof of the results in Section \ref{sec:thm}.
The roadmap to this section is as follows.
We start by introducing a useful technical result in Appendix \ref{apdx:useful.result}.
Appendix \ref{apdx:assumption.fnn} specifies detailed structural assumptions on the feed-forward neural networks and the loss functions.
The proof of Lemma \ref{lm:psi.consistency} is given in Appendix \ref{apdx:pf.lemma.2}.
Appendix \ref{apdx:pf.thm.3} provides the proof of Theorem \ref{thm:main}.
Finally, Corollary \ref{corr:credible.set} is proved in Appendix \ref{apdx:pf.corr.4}.

\subsection{A Useful Result in the Literature}\label{apdx:useful.result}
In the upcoming analysis, the following lemma (see, e.g. Lemma B.1 of \cite{chernozhukov2017monge}) may be useful.
It verifies the equivalence between uniform convergence and continuous convergence.
\begin{lemma}\label{lm:unif.conv.cont.conv}[Equivalent condition of uniform convergence]
    Let $\mathbb{D}$ and $\mathbb{E}$ be complete separable metric spaces, with $\mathbb{D}$ compact.
    Suppose $f:\mathbb{D}\rightarrow\mathbb{E}$ is continuous.
    Then a sequence of functions $f_N:\mathbb{D}\rightarrow\mathbb{E}$ converges to $f$ uniformly if and only if, for any convergent sequence $x_N \rightarrow x$ in $\mathbb{D}$, we have that $f_N(x_N) \rightarrow f(x)$.
\end{lemma}
The proof of Lemma \ref{lm:unif.conv.cont.conv} is given in \cite{rockafellar2009variational}.

\subsection{Structural Assumptions on Neural Networks}\label{apdx:assumption.fnn}
We assume the ReLU activation function of the networks (though the results can be extended to piecewise linear activation functions at no notational cost; see \cite{farrell2021deep}).
Generally, to apply the consistency results in \cite{farrell2021deep} to a function $g$ that is learned by a feed-forward neural network by minimizing the empirical loss function $l(g\mid X,\theta,U)$, it is required to assume that the true underlying function $g_*$ minimizes the expectation of its loss function, i.e. $g_* = \argmin_g \E l(g \mid X,\theta,U)$.
For our method in Section \ref{sec:method} specifically, under the affinity assumption \eqref{eq:affinity.assumption}, the true functions $\varphi, b, f$ (denoted with asterisk in this section) are assumed to satisfy
\[
(\varphi_*, b_*, f_*) = \argmin_{(\varphi, b,f)}\E \mathcal{L}_1(\varphi, b,f \mid \bm{X}, \bm{\theta}, \bm{U}),
\]
where the loss function $\mathcal{L}_1$ is given in \eqref{eq:sun_objective}.

In general, we assume any loss function $l(g \mid X,\theta,U)$ for the learning procedure of any function $g$ (in our case, $\varphi$, $b$ and $f$) to be Lipschitz in $g$, i.e.
\begin{align}\label{eq:loss.assumption.1}
    \begin{split}
        |l(g_1\mid X,\theta,U) &- l(g_2\mid X,\theta,U)| \leq\\
        &C_l |g_1(X,\theta,U) - g_2(X,\theta,U)|
    \end{split}
\end{align}
for any $g_1, g_2$ in the same class as function $g$.
Moreover, $g$ obeys a curvature condition around the true function $g_*$,
\begin{align}\label{eq:loss.assumption.2}
    \begin{split}
        c_1 \E[(g-g_*)^2] \leq \E[l(g\mid X,\theta,U)] &- \E[l(g_*\mid X,\theta,U)]\\
        &\leq c_2 \E[(g-g_*)^2].
    \end{split}
\end{align}
For our method, we assume the loss function $\mathcal{L}_1$ to satisfy \eqref{eq:loss.assumption.1} and \eqref{eq:loss.assumption.2} with respect to $(\varphi, b, f)$.

For a user-chosen architecture $\mathcal{F}_{\text{DNN}}$, the estimate of a function $g$ is computed using generated data flow $\{X_i, \theta_i, U_i\}_{i=1}^N$ by solving
\[
\hat{g}_N := \argmin_{g\in\mathcal{F}_{\text{DNN}}, \|g\|_\infty \leq 2M} \sum_{i=1}^N l(g\mid X_i, \theta_i, U_i).
\]
Our optimization problem \eqref{eq:sun_objective} falls into this category under Condition \eqref{cond:farrell.1} when $N$ is large enough.
Note that in the definition of $\mathcal{L}_1$, the term in the $i$-th summand, $\max_{j\in\{1,\cdots,N\}} \left\{U_j^\top \theta_i - \varphi(U_j) - b(U_j)^\top f(X_i) \right\}$ is essentially a finite-sample estimate to $\sup_{u\in\mathcal{U}} \left\{u^\top \theta_i - \varphi(u) - b(u)^\top f(X_i) \right\}$.
The latter is a function of $\theta_i$ and $X_i$, therefore each summand of $\mathcal{L}_1$ is only related to $X_i, \theta_i$ as long as $N$ is large enough.

\subsection{Proof of Lemma \ref{lm:psi.consistency}}\label{apdx:pf.lemma.2}
    By triangle inequality, $|\hat{\psi}_N(u,x) - \psi(u,x)|$ is upper bounded by
    \begin{align*}
        |\hat{\varphi}_N(u) - \varphi(u)| + \sum_{k=1}^q |\hat{b}_{N,k}(u)\hat{f}_{N,k}(x) - b_k(u)f_k(x)|.
    \end{align*}
    
    By Theorem 2 of \cite{farrell2021deep}, with probability at least $1-e^{-\gamma}$ over the random sample drawn in data generation process,
    \begin{align*}
        \E_U (\hat{\varphi}_N(U) - \varphi(U))^2 \leq \Phi_\varphi(N,\gamma).
    \end{align*}
    Similarly, for any $1\leq k\leq q$, with probability at least $1-e^{-\gamma}$,
    \begin{align*}
        \E_U (\hat{b}_{N,k}(U) - b_k(U))^2 \leq \Phi_b(N,\gamma),
    \end{align*}
    and with probability at least $1-e^{-\gamma}$,
    \begin{align*}
        \E_{X} (\hat{f}_{N,k}(X) - b_k(X))^2 &\leq \Phi_f(N,\gamma).
    \end{align*}
    Here, the finite sample rates are defined as follows.
    \begin{align}\label{eq:farrell.rate}
        \begin{split}
            &\Phi_\varphi(N,\gamma) =\\
            &C_\varphi\left(\frac{W_\varphi L_\varphi\log W_\varphi}{N}\log N + \frac{\log\log N + \gamma}{N} + (\epsilon_{\text{DNN}}^\varphi)^2\right);\\
            &\Phi_b(N,\gamma) =\\
            &C_b\left(\frac{W_b L_b\log W_b}{N}\log N + \frac{\log\log N + \gamma}{N} + (\epsilon_{\text{DNN}}^b)^2\right);\\
            &\Phi_f(N,\gamma) =\\
            &C_f\left(\frac{W_f L_f\log W_f}{N}\log N + \frac{\log\log N + \gamma}{N} + (\epsilon_{\text{DNN}}^f)^2\right),
        \end{split}
    \end{align}
    where $C_\varphi, C_b, C_f$ are positive constants that are independent of $N$, $W_\varphi, W_b, W_f$ are the maximum widths of the respective neural networks at a given layer, and $L_\varphi, L_b, L_f$ are the maximum depths of the respective neural networks.
    By Cauchy-Schwarz inequality,
    \begin{align*}
        &(\hat{\psi}_N(u,x) - \psi(u,x))^2\\
        \leq& (2q+1)(\hat{\varphi}_N(u) - \varphi(u))^2\\
        &+ (2q+1)\sum_{k=1}^q \hat{f}_{N,k}^2(x)(\hat{b}_{N,k}(u) - b_k(u))^2\\
        &+ (2q+1)\sum_{k=1}^q b_k^2(u)(\hat{f}_{N,k}(x) - f_k(x))^2.
    \end{align*}
    Therefore, with probability no less than $1-3qe^{-\gamma}$, we have the following $L^2$ convergence guarantee of the estimated potential function:
    \begin{align}\label{eq:L2.convergence}
        \begin{split}
            &\E_U \E_{X}( \hat{\psi}_N (U,X) - \psi(U,X))^2\\
            &\leq 3q\Phi_\varphi(N,\gamma) + 12q^2 M^2\Phi_b(N,\gamma) + 3q^2 M^2 \Phi_f(N,\gamma).
        \end{split}
    \end{align}
    We denote the event $\mathcal{A}_N$ hereafter, under which \eqref{eq:L2.convergence} holds.
    By Markov's inequality, this $L^2$ convergence immediately implies convergence in probability.
    
    Meanwhile, under Condition \ref{cond:farrell.1}, the estimated function $\{\hat{\psi}_N\}_{N\in\N}$ also satisfies uniform boundedness and uniform equicontinuity on their domain $\mathcal{U}\times \mathcal{X}$, which is also compact.
    By Arzel\`a-Ascoli Theorem, there exists a subsequence $\{\hat{\psi}_{N_m}\}_{m\in \N}$ that converges uniformly.
    The continuity of each function $\hat{\psi}_{N_m}$ implies that the uniform limit $\hat{\psi}^\diamond := \lim_{m\rightarrow\infty} \hat{\psi}_{N_m}$ is also continuous.
    On the other hand, under event $\mathcal{A}_N$, $\hat{\psi}_{N_m}$ converges to $\psi$ in probability.
    We can thus extract a further subsequence $\{\hat{\psi}_{N_{m_l}}\}_{l\in \N}$, which also uniformly converges to $\hat{\psi}^\diamond$, such that $\hat{\psi}_{N_{m_l}} \rightarrow \psi$ almost everywhere on $\mathcal{U}\times \mathcal{X}$.
    Therefore $\hat{\psi}^\diamond = \psi$ almost everywhere on $\mathcal{U} \times \mathcal{X}$.
    Since both functions are continuous, $\hat{\psi}^\diamond(u,x) = \psi(u,x)$ holds for all $(u,x)\in\mathcal{U}\times\mathcal{X}$.
    This ensures uniqueness of the uniform limit, which further implies that the sequence $\hat{\psi}_N$ uniformly converges to $\psi$, finishing the proof of the the first assertion under the event $\mathcal{A}_N$.

    For the second assertion, note that the uniform convergence of $\hat{\psi}_N$ implies that for any $\epsilon>0$, there exists $N_0\in\N$ such that for all $N\geq N_0$ and $u\in\mathcal{U}$, $\hat\psi_N(u,x) \leq \psi(u,x) + \epsilon$.
    Therefore,
    \[
    \sup_{u\in\mathcal{U}} \{\theta^\top u - \hat \psi_N(u,x)\} \geq \sup_{u\in\mathcal{U}} \{\theta^\top u - \psi(u,x) - \epsilon\},
    \]
    which is equivalent to $\hat \psi_N^*(\theta,x) \geq \psi^*(\theta,x) - \epsilon$, and
    \[
    \liminf_{N\rightarrow\infty} \hat \psi_N^*(\theta,x) \geq \psi^*(\theta,x) - \epsilon.
    \]
    Similarly, we have $\hat \psi_N^*(\theta,x) \leq \psi^*(\theta,x) + \epsilon$, and
    \[
    \limsup_{N\rightarrow\infty} \hat \psi_N^*(\theta,x) \leq \psi^*(\theta,x) + \epsilon.
    \]
    Since $\epsilon>0$ can be arbitrarily small, we first arrive at pointwise convergence: 
    \begin{equation} \label{eq:conjugate.pointwise.conv}
        \lim_{N\rightarrow\infty} \hat \psi_N^*(\theta,x) = \psi^*(\theta,x).
    \end{equation}

    Given any compact subset $K'\subset \Theta$, we want to show that the convergence is uniform, i.e.
    \[
    \sup_{\theta\in K', x\in \mathcal{X}} |\hat\psi_N^*(\theta,x) - \psi^*(\theta,x)| \rightarrow 0.
    \]
    Consider a sequence in $K'\times\mathcal{X}$, $(\theta_N, X_N) \rightarrow (\theta,X)$.
    Since $\mathcal{U}\times\mathcal{X}$ is compact, the supremum in the definition of $\psi^*$ is attained within $\mathcal{U}\times\mathcal{X}$.
    Therefore, for any $\theta\in\Theta$, there exists $u(\theta)\in\mathcal{U}$ such that
    \[
    \psi^*(\theta,x) = u(\theta)^\top \theta - \psi(u(\theta),x).
    \]
    Similarly, for any $N\in\N$ and $\theta_N\in\Theta$, there exists $u_N(\theta_N)\in\mathcal{U}$ such that
    \[
    \hat \psi_N^*(\theta,x_N) = u_N(\theta_N)^\top \theta_N - \hat \psi_N (u_N(\theta_N),x_N).
    \]
    Note that $u(\theta)$ is the maximizer in the definition of $\psi^*(\theta,x)$.
    Therefore we have the inequality $\psi^*(\theta,x) \geq u_N(\theta_N)^\top\theta - \psi(u_N(\theta_N),x)$, thus the difference $\hat \psi_N^*(\theta_N,x_N) - \psi^*(\theta,x)$ is upper bounded by
    \[
    u_N(\theta_N)^\top(\theta_N-\theta) - [\hat \psi_N(u_N(\theta_N),x_N) - \psi(u_N(\theta_N), x)].
    \]
    The first term converges to zero since $\theta_N\rightarrow\theta$ and that $\mathcal{U}$ is a compact set.
    the two terms in the bracket also converges to zero, by combining uniform convergence of $\hat\psi_N$, i.e.
    \[
    \hat\psi_N(u_N(\theta_N),x_N) - \psi(u_N(\theta_N),x_N) \rightarrow 0,
    \]
    with equicontinuity of $\psi$, i.e.
    \[
    \psi(u_N(\theta_N),x_N) -  \psi(u_N(\theta_N),x)\rightarrow 0.
    \]
    Therefore
    \begin{equation}\label{eq:psi.N.limsup}
        \limsup_{N\rightarrow\infty} [\hat \psi_N^*(\theta_N,x_N) - \psi^*(\theta,x)] \leq 0.
    \end{equation}
    Using the same method, $\psi^*(\theta,x) - \hat \psi_N^*(\theta_N,x_N)$ is upper bounded by
    \[
    u(\theta)^\top(\theta-\theta_N) + [\psi(u(\theta),x_N) - \hat\psi_N(u(\theta),x)],
    \]
    which also converges to zero, yielding
    \begin{equation}\label{eq:psi.N.liminf}
        \liminf_{N\rightarrow\infty} [\hat \psi_N^*(\theta_N,x_N) - \psi^*(\theta,x)] \geq 0.
    \end{equation}
    Combining \eqref{eq:psi.N.limsup} and \eqref{eq:psi.N.liminf}, we arrive at
    \begin{equation}\label{eq:psi.N.lim}
        \hat \psi_N^*(\theta_N,x_N) \rightarrow \psi^*(\theta,x) \text{ for all }(\theta_N,x_N)\rightarrow(\theta,x).
    \end{equation}
    The proof of the second assertion under the event $\mathcal{A}_N$ is thus finished by Lemma \ref{lm:unif.conv.cont.conv}.

    Finally, the almost sure argument is proved by noting that $\mathbb{P}(\mathcal{A}_N^c) \leq 3qe^{-\gamma}$.
    By taking $\gamma = 2\log N$, the rates in \eqref{eq:farrell.rate} still converge to zero as $N\rightarrow\infty$.
    Additionally, we have
    \[
    \sum_{N=1}^\infty \mathbb{P}(\mathcal{A}_N^c) \leq \sum_{N=1}^\infty \frac{3q}{N^2} <\infty.
    \]
    By Borel-Cantelli Lemma, this implies event $\mathcal{A}_N$ occurs indefinitely often.

\subsection{Proof of Theorem \ref{thm:main}}\label{apdx:pf.thm.3}
Recall that
\begin{align*}
    \hat Q^N_{\theta\mid X=x}(u) &= \argsup_{\theta\in\Theta} [\theta^\top u - \hat\psi^*_N(\theta,x)],\\
    Q_{\theta\mid X=x}(u) &= \argsup_{\theta\in\Theta} [\theta^\top u - \psi^*(\theta,x)].
\end{align*}
Define the extension map $g_{x,u}^N(\theta)= \theta^\top u - \hat\psi^*_N(\theta,x)$ if $\theta\in\Theta$, and $-\infty$ if $\theta\notin \theta$.
Another extension map $g_{x,u}(\theta)$ can be defined analogously with $\psi^*(\theta,x)$.
Note that this is a concave mapping from $\R^d$ to $\R$.
By Lemma \ref{lm:psi.consistency}, $\hat\psi^*_N$ is uniformly consistent over any compact set $K'\subset\Theta$.
Under event $\mathcal{A}_N$, for any sequence $\{u_N\}$ such that $u_N \rightarrow u\in K_U$, a compact set of $\mathcal{U}_0$, and any $x\in\mathcal{X}_0$,
\[
g^N_{x,u_N}(\theta) \rightarrow g_{x,u}(\theta)
\]
for all $\theta\in\text{int }\Theta$, a dense subset of $\Theta$.
By Lemma B.2 of \cite{chernozhukov2017monge}, we have the consistency of the argsup,
\[
\argsup_{\theta\in\Theta} g_{x,u_N}^N(\theta) \rightarrow \argsup_{\theta\in\Theta} g_{x,u}(\theta),
\]
which is equivalent to
\[
\hat Q^N_{\theta\mid X=x}(u_N) \rightarrow Q_{\theta\mid X=x}(u).
\]
Since the sequence $u_N\rightarrow u$ is arbitrarily taken over the compact subset $K$, by Lemma \ref{lm:unif.conv.cont.conv}, we have
\[
\hat Q^N_{\theta\mid X=x}(u) \rightarrow Q_{\theta\mid X=x}(u) \text{ uniformly over }K_U.
\]
under event $\mathcal{A}_N$.
The first assertion is then proved by letting $\gamma = 2\log N$ and applying Borel-Cantelli Lemma, like in the proof of Lemma \ref{lm:psi.consistency}.

Since our method solves the Monge-Kantorovich problem \eqref{eq:monge}, the optimal transport mapping ensure the following identity of 2-Wasserstein distance,
\begin{align}\label{eq:W2.identity}
    \begin{split}
        W_2^2&(\hat{\pi}^N(\theta \mid X=x), \pi(\theta\mid X=x))\\
        &= \int_\mathcal{U}\|\hat Q^N_{\theta\mid X=x}(u) - Q_{\theta\mid X=x}(u)\|^2 \diff F_U(u).
    \end{split}
\end{align}
Note that our source distribution $F_U$ features a density that is proportional to $r^{-d+1}$, which is positive everywhere on the entire unit ball $\mathcal{U} = S^d(1)$ and upper bounded except on the center.
Since the choice of compact subset $K_U \subset \mathcal{U}_0 = \text{int }S^d(1)$ is arbitrary, it is possible to find a $K_U$ that contains the center such that its Lebesgue measure is arbitrarily close to that of $\mathcal{U}$.
In that case, for any $\epsilon>0$ there exists such a compact set $K_U$ that
\[
F_U(\mathcal{U}\setminus K_U) < \frac{\epsilon}{4dM^2}.
\]
In light of this, the right-hand side of \eqref{eq:W2.identity} can be upper bounded by splitting the integral,
\begin{align*}
    &\int_{K_U} \|\hat Q^N_{\theta\mid X=x}(u) - Q_{\theta\mid X=x}(u)\|^2 \diff F_U(u)\\
    &+ \int_{\mathcal{U}\setminus K_U} \|\hat Q^N_{\theta\mid X=x}(u) - Q_{\theta\mid X=x}(u)\|^2 \diff F_U(u).
\end{align*}
The first term converges to zero almost surely.
Since the parameter space $\Theta\subset[-M,M]^d$, the second term is upper bounded by $4dM^2 F_U(\mathcal{U}\setminus K_U) <\epsilon$.
This proves the convergence of $W_2(\hat{\pi}^N(\theta \mid X=x), \pi(\theta\mid X=x))$ to zero given any fixed $x\in K_X\subset\mathcal{X}$ under the event $\mathcal{A_N}$, which occurs almost surely.

\subsection{Proof of Corollary \ref{corr:credible.set}}\label{apdx:pf.corr.4}
Given any $\tau\in(0,1)$, the ball $S^d(\tau)$ is a closed subset in the interior of $\mathcal{U} = S^d(1)$.
Due to the uniform convergence result of vector quantiles given in the first assertion of Theorem \ref{thm:main}, almost surely we have
\begin{equation}\label{eq:unif.conv.ball}
\sup_{x\in K_X}\sup_{u\in S^d(\tau)} \|\hat Q^N_{\theta \mid X=x}(u) - Q_{\theta \mid X=x}(u)\| \rightarrow 0,
\end{equation}
i.e. uniform convergence over the ball $S^d(\tau)$.

We next prove the Hausdorff distance between the credible set $\hat C^N_\tau (\theta \mid X=x)$ with the oracle set $C_\tau (\theta \mid X=x)$.
For any $\theta\in C_\tau (\theta \mid X=x)$, since the ball $S^d(\tau)$ is compact, there exists $u_0\in S^d(\tau)$ such that $\theta = Q_{\theta\mid X=x}(u_0)$.
By uniform convergence \eqref{eq:unif.conv.ball},
\[
\sup_{x\in K_X}\| \hat Q^N_{\theta \mid X=x}(u_0) - \theta \| \rightarrow 0.
\]
Since $u_0 \in S^d(\tau)$, by definition of the oracle set \eqref{eq:oracle.set}, we have $\hat \theta^N := \hat Q^N_{\theta\mid X=x}(u_0) \in \hat C^N_\tau (\theta \mid X=x)$.
This means for any $\theta\in C_\tau (\theta \mid X=x)$, there exists $\hat \theta^N \in \hat C^N_\tau (\theta \mid X=x)$ such that $\|\hat \theta^N - \theta\|\rightarrow 0$. Thus
\[
\sup_{\theta\in C_\tau (\theta \mid X=x)} \inf_{\hat\theta^N\in\hat C^N_\tau (\theta \mid X=x)} \|\hat\theta^N - \theta \| \rightarrow 0.
\]
Conversely, we can prove that for any $\hat \theta^N \in \hat C^N_\tau (\theta \mid X=x)$, there exists $\theta\in C_\tau (\theta \mid X=x)$ such that $\|\hat \theta^N - \theta\|\rightarrow 0$, and thus
\[
\sup_{\hat\theta^N\in\hat C^N_\tau (\theta \mid X=x)} \inf_{\theta\in C_\tau (\theta \mid X=x)} \|\hat\theta^N - \theta \| \rightarrow 0.
\]
Since all convergences are uniform over $x\in K_X$, we conclude that
\[
\sup_{x\in K_X} d_H(\hat C^N_\tau (\theta \mid X=x), C_\tau (\theta \mid X=x)) \rightarrow 0.
\]

\section{Autoregressive Quantile Learning}\label{sec:autoreg} Another way to use quantile modeling for multi-dimensional data is through auto-regressive modeling. This idea was first proposed by \cite{koenker2006quantile} and later adopted in the machine learning literature for generative modeling \citep{ostrovski2018autoregressive} and time series forecasting \citep{gouttes2021probabilistic}. The key idea is to define the joint quantile by using one-dimensional conditional quantiles as \( F_X(x) = P(X_1 \leq x_1, \ldots, X_n \leq x_n) = \prod_{i=1}^n F_{X_i \mid X_{i-1}, \ldots, X_1}(x_i) \). Now we can define \( F_X^{-1}(\tau_1, \ldots, \tau_n) = (F_{X_1}^{-1}(\tau_1), \ldots, F^{-1}_{X_n \mid X_{n-1}, \ldots}(\tau_n)) \). Each one-dimensional quantile can be optimized by minimizing the pinball loss integrated over all quantile levels $\tau\in[0,1]$, and they are often designed with RNNs to encode the conditions. Due to the sequential conditional structure, this approach has received attention in modeling time-dependent latent variables, such as implied volatility. 

\section{More Results for the Gaussian Example}\label{sec:gauss}
\paragraph{More details for the Gaussian example experiment in Section \ref{sec:nu}} The model allow us to simulate a pair of parameter and data (of $n$ observations) as $(X, (\mu,\sigma^2))$ for $X\in \R^n$. Due to the conjugate prior, the true posterior samples are available. For the observed data, given the number of observations $n\geq 1$, we set $X = [x,...,x]\in \R^n$ for a few manually chosen $x$ values. The feature map is chosen as a DeepSet with the output dimension 2. We trained all our network for 150 epochs, with 1 epoch as 100 iterations with mini-batch size 128. We optimize by the Adam optimizer \citep{diederik2014adam} with default hyperparameter settings, where the learning rate is 0.01 and is reduced by $\times 0.99$ for each epoch. We found that a good initial random initialization is important to learn a meaningful quantile map. We run 10 different random trials and then pick up the model having the lowest loss function. 

\begin{figure}[!h]
    \centering
    \includegraphics[width=\linewidth]{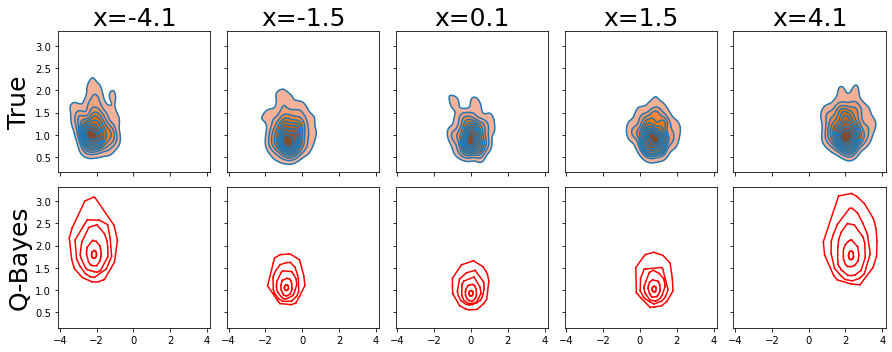}
    \caption{The credible set for the Gaussian example with varying $x$ with $n=2$. The subplots show the alignment on the first axis (x-axis) between the true posterior sample and the generated samples.}
    \label{fig:gaussian_n_inc}
\end{figure}

\paragraph{The case of $n=2$.} Here, we consider the case when ample size $n=2$. To obtain the credible set of probability $\tau$, we sample $U\sim \tau F_U$, which is equivalent to the source distribution restricted on radius $\tau$, and apply the trained quantile sampler. The boundary of the credible set is obtained by applying a contour plot on the sampled credible sets. The results are shown in Figure \ref{fig:gaussian_n_inc}. We can see very successful posterior sampling, and the credible sets show the nested structure.

\begin{figure}[!h]
    \centering
    \includegraphics[width=\linewidth]{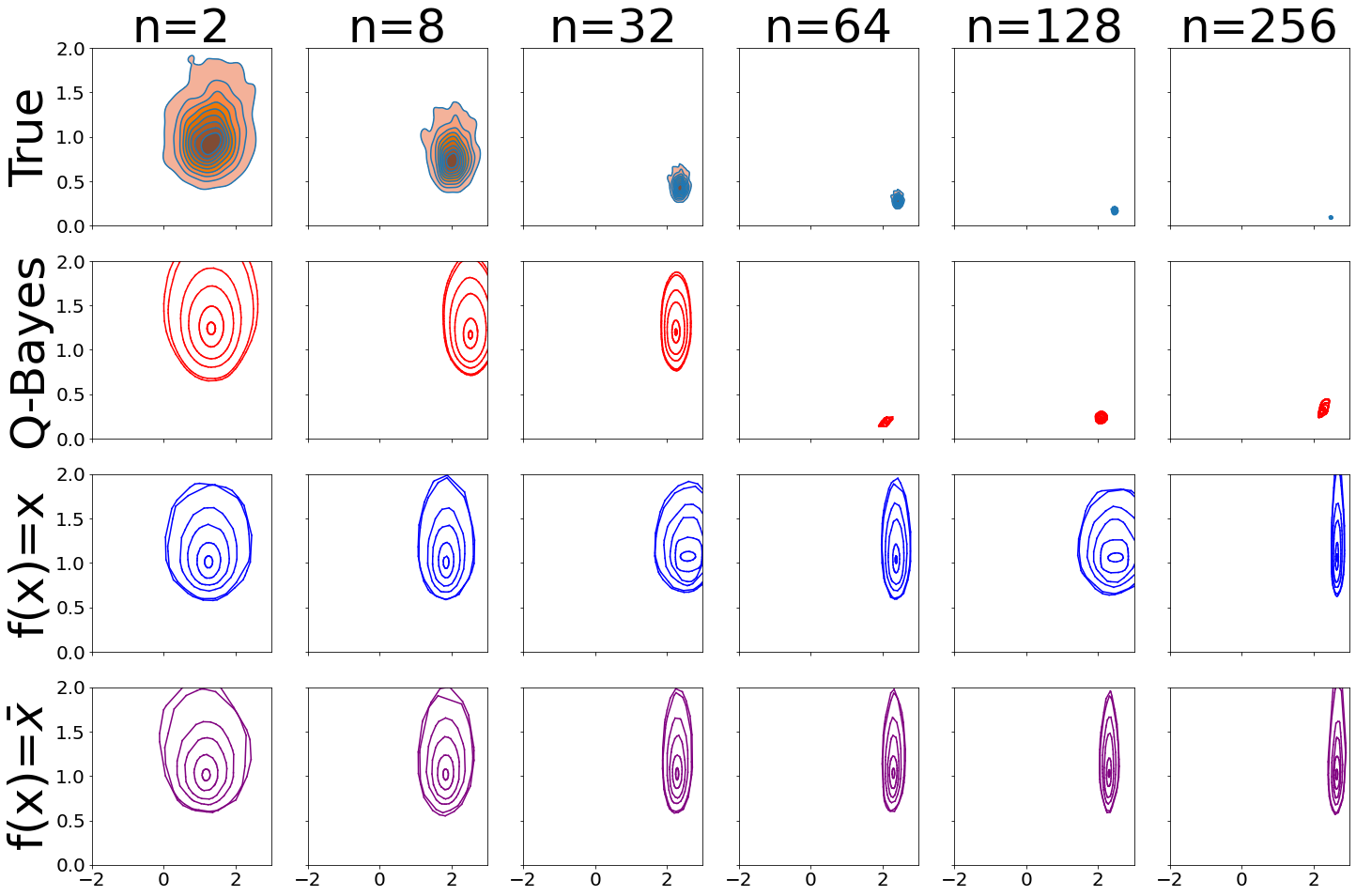}
    \caption{The credible set for the Gaussian example with increasing $n$ for a relatively large $x$ ($x=2.5$). From $n=16$, the quality decreases.}
    \label{fig:bad}
\end{figure}

\paragraph{The case of increasing $n$.} We already highlighted in Section \ref{sec:nu} that the support shrinkage is clearly observed with an increasing $n$ for $x$ relatively near the origin. A similar phenomenon is observed also for a relatively large $x$ as in Figure \ref{fig:bad} albeit some performance decrease probably due to lack of observed data during training. 

\begin{figure}
    \centering
    \includegraphics[width=\linewidth]{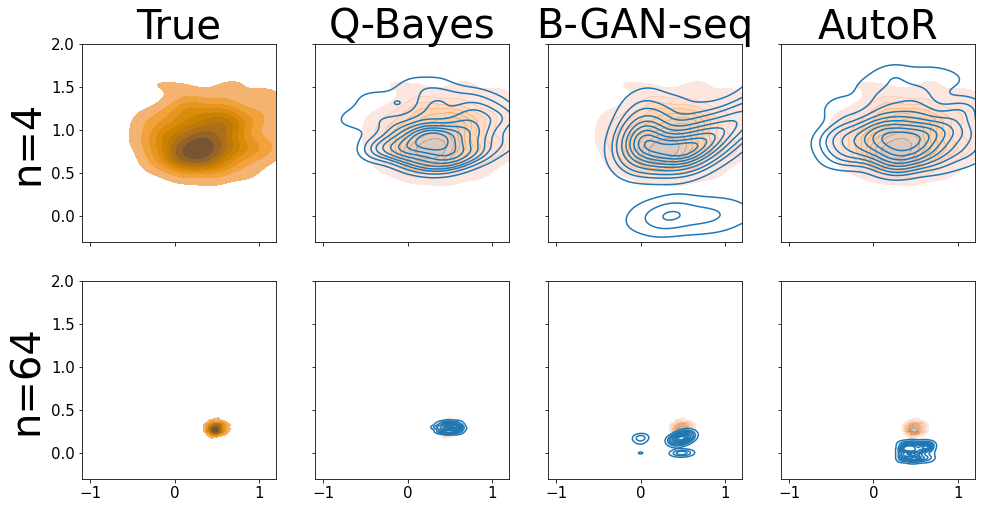}
    \caption{Comparison between Auto Regression (AutoR), Sequential Adversarial Bayesian Simulation (B-GAN-seq), and the Vector Quantile (Brenier) methods for $n=4$ and $n=64$. The contour lines in blue represent the KDE approximation.}\label{fig:gaussian_eamples}
\end{figure}

\begin{figure}[!h]
    \centering
    \includegraphics[width=\linewidth]{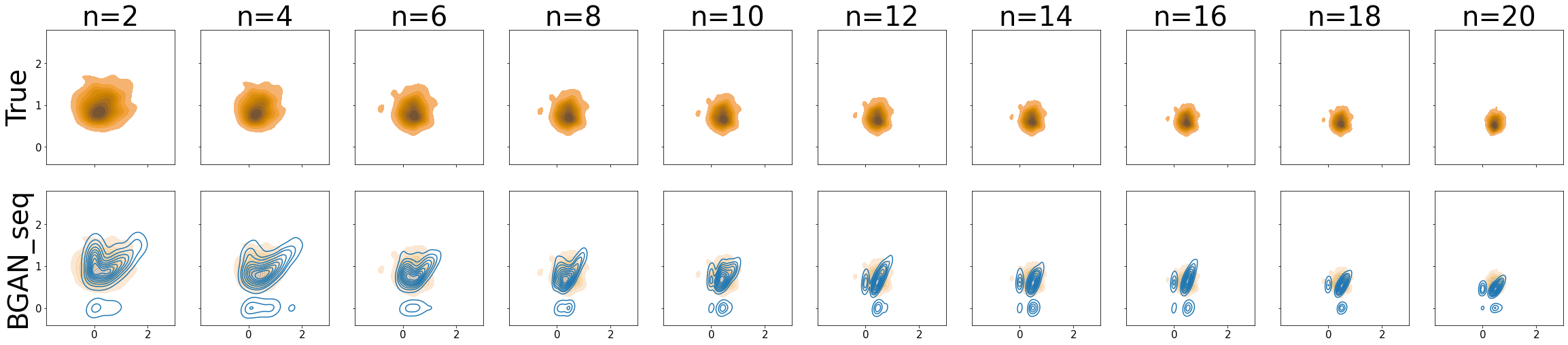}
    \caption{The sequential B-GAN method for increasing $n$. The network is trained sequentially for every additional two data points. The contour lines in blue represent the KDE approximation.}
    \label{fig:gaussian_eamples_abc}
\end{figure}

\begin{figure}[!h]
    \centering
    \includegraphics[width=\linewidth]{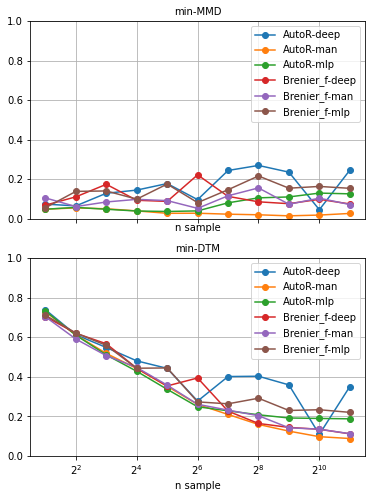}
    \caption{The minimum MMD and DTM (among 10 random runs) for increasing $n$.}
    \label{fig:gaussian_eamples3}
\end{figure}

\paragraph{Comparison with other methods.}
We train a sequential B-GAN (B-GAN-seq, \cite{wang2022adversarial}) and the Autoregressive method from Section \ref{sec:autoreg} (AutoR) using an off-the-shelf sampling mechanism and compare the results in Figure \ref{fig:gaussian_eamples}. We believe that the scalability achieved by adopting DeepSets has contributed to the improvement in our algorithm. To examine the effect of DeepSet more clearly, we train our method with three different feature networks $f$: $f$ as a DeepSet (Q-Bayes-deep), a fully connected network (Q-Bayes-mlp), and the mean and standard deviation as manually chosen summary statistics (Q-Bayes-man).
Performance is measured by the Maximum Mean Discrepancy (MMD) between the empirical distribution of the true posterior and the generated samples. We also measure Distance to Measure (DTM), which is the average distance between the true parameters and the generated posterior samples conditioned on the data from the true parameters. Specifically, given a trained posterior-generating model $Q(\theta\mid X)$, DTM is defined by $\frac{1}{3\times 10^4}\sum_{j=1}^{100} \sum_{i=1}^{300} |\theta_j^*-\theta_{ji}|$, where $\theta_j^*\sim\pi(\theta)$ and $\theta_{ji}\sim Q(\theta\mid X_j)$ for $X_j\sim L(X\mid \theta^*_j).$ To account for the variability introduced by the randomness of each run, we perform 10 runs for each method and select the model with the best performance measure.
As shown in Figure \ref{fig:gaussian_eamples3}, the fully connected network (Q-Bayes-mlp) does not perform as well as the DeepSet (Q-Bayes-deep) as $n$ increases. A similar improvement might be possible for B-GAN with the development of a variation that incorporates a learnable feature network to enable the use of DeepSets. Currently, B-GAN handles increasing $n$ by sequentially applying B-GAN, using a previously learned posterior sampler as a new prior sampler. (For the support shrinkage of B-GAN-seq, see Figure \ref{fig:gaussian_eamples_abc}.)
We also train the autoregressive model with the three feature maps (AutoR-deep, AutoR-mlp, AutoR-man), as shown in Figure \ref{fig:gaussian_eamples3}. Note that in this Gaussian setting, the summary statistics are sufficient summary statistics. The Autoregressive method with the appropriate summary statistics demonstrates the best performance. On the other hand, with other feature $f$ networks, it falls short of our DeepSet-based method as $n$ increases.

{Note that our method is not the only method that can demonstrate this effect; The sequential B-GAN (B-GAN-seq) and auto-regressive method (AutoR) show support shrinkage. For AutoR, as shown in Figure \ref{fig:gaussian_eamples3}, its performance is only competitive when the true summary statistics are given (the orange line). As a comparison, our proposed method shows about the same performance even when the summary statistics are unknown and learned automatically (the red line). For B-GAN, it had to be adapted in these comparisons to exhibit support shrinkage. While the quality of the samples from B-GAN-seq is very competitive, it trains a sampler for a fixed $X_0$, not a generic realization of the data. }

\section{Network Architecture}

In all our experiments, {the DeepSet feature network is implemented with input dimension $d_X$, a set feature space dimension $q_1$, a hidden layer width of 16, and two layers of width 16 in the transformation network. We use the LSTM network design by \cite{sun2022conditional}.  For the LSTM feature network, the architecture consists of a single LSTM layer with a hidden state size of 512 and an input size of \( d_X \). The final output of the sequence is mapped to a fixed dimensionality \( q_2 \) using a fully connected layer. In what follows, we provide the actual code to implement them.}

\subsection*{Multilayer Perceptron (MLP)}
\begin{lstlisting}[style=pythonstyle]
import torch
import torch.nn as nn

class MLP(nn.Module):
    def __init__(self, device="cuda", dim=2, z_dim=1, 
                 leaky=0.1, factor=64, n_layers=2, 
                 dropout=0, positive=False):
        super().__init__()
        self.non_linear = nn.LeakyReLU(leaky) if leaky > 0 else nn.ReLU()
        self.dropout = nn.Dropout(dropout)
        self.layers = nn.ModuleList([nn.Linear(dim, factor)])
        for _ in range(n_layers):
            self.layers.append(nn.Linear(factor, factor))
        self.layers.append(nn.Linear(factor, z_dim))
        self.positive = positive
        self.to(device)

    def forward(self, x):
        h = x
        for i, layer in enumerate(self.layers):
            h = layer(h)
            if i < len(self.layers) - 1:  # Apply non-linearity except last layer
                h = self.non_linear(h)
                h = self.dropout(h)
        return h.abs() if self.positive else h
\end{lstlisting}

\subsection*{The LSTM Feature Network}
\begin{lstlisting}[style=pythonstyle]
class BiRNN(nn.Module):
    def __init__(self, input_size, hidden_size, num_layers, xdim, bn_last=True, device="cuda"):
        super(BiRNN, self).__init__()
        self.hidden_size = hidden_size
        self.num_layers = num_layers
        self.bn_last = bn_last
        self.lstm = nn.LSTM(input_size, hidden_size, num_layers, batch_first=True, bidirectional=False)
        self.fc = nn.Linear(hidden_size, xdim)
        self.norm = nn.BatchNorm1d(xdim, momentum=1.0, affine=False)
        self.device = device
        self.to(device)

    def forward(self, x):
        # Set initial states
        h0 = torch.zeros(self.num_layers, x.size(0), self.hidden_size).to(self.device)
        c0 = torch.zeros(self.num_layers, x.size(0), self.hidden_size).to(self.device)
        # Forward propagate LSTM
        out, _ = self.lstm(x, (h0, c0))
        # Decode the hidden state of the last time step
        out = self.fc(out[:, -1, :])
        if self.bn_last:
            return self.norm(out)
        return out
\end{lstlisting}

\subsection*{The DeepSet Feature Network}
\begin{lstlisting}[style=pythonstyle]
class DeepSets(nn.Module):
    def __init__(self, dim_x, dim_ss, factor=16, 
                 num_layers=2, device="cuda", 
                 bn_last=True):
        super(DeepSets, self).__init__()
        self.common_feature_net = MLP(device=device,dim=dim_x,z_dim=dim_ss,dropout=0, factor=64, n_layers=3)
        self.next_net = MLP(device=device, dim=dim_ss, z_dim=dim_ss, factor=factor, n_layers=num_layers)
        self.to(device)
        self.device = device
        self.bn_last = bn_last
        self.norm = nn.BatchNorm1d(dim_ss, momentum=1.0, affine=False)

    def forward(self, x):
        shape = x.shape
        assert len(shape) == 3
        phi = self.common_feature_net(x.view(-1, shape[-1])).view(x.shape[0], x.shape[1], -1).mean(1)
        out = self.next_net(phi)
        if self.bn_last:
            return self.norm(out)
        return out
\end{lstlisting}

\section{Comparison on the Brock Hommes model}\label{sec:table_detail}
{

{We compare our method with the standard ABC (rejection ABC) and sequential Monte Carlo ABC (SMC-ABC, \cite{sisson2007sequential}), which is designed to enhance the efficiency of standard ABC methods. For ABC, we allocated a total budget of $10^8$ proposals, a generous allocation commonly used for complex ABC simulations \citep{raynal2019abc}. Since summary statistics are unknown, we adopted a naive version of ABC by using the entire data vector as the statistic and employing the $L_2$ distance as the metric for thresholding.
To ensure at least 1,000 samples were selected, the threshold was set to 22, resulting in 1,153 proposals accepted out of the total of one hundred million proposals. For SMC-ABC, we use the implementation provided by \cite{o2024tree}. For numerical performance metrics, we use DTM and the distance from the posterior mean (DPM), along with the computational time.



\begin{table}
    \centering
    \begin{tabular}{c|c|c|c|c}
         Method & Ours & \makecell{Rejection \\ ABC} & \makecell{SMC\\ABC} & AutoR \\\hline
         DTM    & 0.177 & 0.521 & 0.513 & 0.179 \\\hline
         DPM    & 0.159 & 0.432 & 0.441 & 0.195 \\\hline
         Time   & 37 mins & 29 mins & 41 mins & 27 mins \\
    \end{tabular}
    \caption{Comparisons on the Brock-Hommes model}
    \label{tb:table1}
\end{table}

The results are given in Table \ref{tb:table1}. Without a big increase in computational cost, our method produces a qualitative improvement in the posterior samples. It is important to emphasize that ABC sampling must be performed separately for each new observed dataset, making it computationally expensive collectively. On the contrary, our method learns a mapping. Once such a mapping is trained, it can be applied to any observed dataset at a very low computational cost without the need for retraining. In aggregate, this significantly improves efficiency in practice.}

\end{document}